\documentclass[11pt]{article}
\usepackage{amsmath,amssymb,amsfonts,amsthm,epsfig}
\usepackage[dvipsnames]{xcolor}
\usepackage{bm,xspace}
\usepackage{tcolorbox}
\usepackage{cancel}
\usepackage{fullpage}
\usepackage{liyang}
\usepackage{framed}
\usepackage{verbatim}
\usepackage{enumitem}
\usepackage{array}
\usepackage{multirow}
\usepackage{afterpage}
\usepackage{mathrsfs}
\usepackage{pifont} 
\usepackage{chngpage}
\usepackage[normalem]{ulem}
\usepackage{boxedminipage}
\usepackage{caption}
\usepackage{subcaption}
\usepackage{bold-extra}
\usepackage{forest}
\usepackage{etoolbox}
\usepackage{thm-restate}
\usepackage{microtype}

\usepackage{algorithm}
\usepackage{algorithmicx}
\usepackage{algpseudocode}

\usepackage{tikz}
\usetikzlibrary{calc,through,backgrounds,decorations.pathreplacing, calligraphy,arrows.meta}
\usetikzlibrary{positioning,chains,fit,shapes}
\usetikzlibrary{patterns,patterns.meta}

\usepackage{tikz-cd}

\usepackage{pgfplots}
\pgfplotsset{width=8cm,compat=newest}

\def\colorful{1}

\ifnum\colorful=1

\fi
\ifnum\colorful=0

\fi

\newcommand{\pparagraph}[1]{\bigskip \noindent {\bf {#1}}}

\newcommand{\BQP}{\mathsf{BQP}}
\newcommand{\PromiseBQP}{\mathsf{PromiseBQP}}
\newcommand{\BPP}{\mathsf{BPP}}
\newcommand{\PromiseBPP}{\mathsf{PromiseBPP}}

\renewcommand{\P}{\mathsf{P}}
\newcommand{\PromiseHeurBPP}{\mathsf{PromiseHeurBPP}}

\newcommand{\row}{\mathrm{row}}
\newcommand{\minimal}{\mathrm{minimal}}
\newcommand{\errors}{\mathrm{errors}}

\Crefname{theorem}{Theorem}{Theorems}
\Crefname{thmenumi}{Theorem}{Theorems}
\AtBeginEnvironment{theorem}{%
    \crefalias{enumi}{thmenumi}%
    \setlist[enumerate,1]{
        label={\textit{(\roman*)}},
        ref={\thetheorem(\roman*)}
    }
}

\newcommand{\Infone}{\Inf^{(1)}}

\newcommand{\Lone}[1]{\lVert #1\rVert_1}
\newcommand{\Abar}{\overline{A}}

\newcommand{\FC}{\mathrm{FC}}
\newcommand{\calA}{\mathcal{A}}

\newcommand{\NP}{\mathsf{NP}}

\newcommand{\AvgP}{\mathsf{AvgP}}

\newcommand{\cube}{\mathrm{cube}}

\newtcolorbox{resultbox}{colback=white, colframe=black!75, boxrule=0.8pt, arc=4pt,
  left=6pt, right=6pt, top=2pt, bottom=2pt, before skip=10pt, after skip=10pt}

\begin{document}

\title{

Proof of the Aaronson--Ambainis Conjecture: \\ 
An Exposition\vspace{15pt}
}

\author{ 
Guy Blanc \vspace{6pt} \\ 
\hspace{-7pt} {\sl MIT} \and 
Jordan Docter \vspace{6pt} \\ 
\hspace{-10pt} { {\sl Stanford}} \and 
 Carmen Strassle \vspace{6pt} \\
 \hspace{-5pt} { {\sl Stanford}} \vspace{15pt}
 \and 
Li-Yang Tan \vspace{6pt}  \\
\hspace{-10pt} {{\sl Stanford}}
}

\date{\small{\today}}

\pagenumbering{gobble}
 \maketitle

\begin{abstract}
We give an exposition of a recent proof of the Aaronson--Ambainis conjecture, which states that every degree-$d$ polynomial $p : \zo^n \to [0,1]$ has a variable of influence at least $\poly(\Var(p)/d)$. This implies that  quantum query algorithms for decision problems can be efficiently simulated classically on most inputs.  It also yields  a near-equivalence between the status of the $\mathsf{BPP}$ vs.~$\mathsf{BQP}$ problem in the random oracle and unrelativized worlds.

We organize the proof around four main steps and discuss how each connects to the literature on the analysis of boolean functions and probabilistic combinatorics. To bring out the key ideas, we do not optimize parameters, and our presentation yields weaker exponents than the original.

\bigskip 

\bigskip

\bigskip 

\noindent {\bf Authors' note.} The proof was discovered by David Turturean using a research system developed at Physical Superintelligence (PSI), in combination with a non-public Anthropic model.  The proof was then shared with the authors, who worked to understand  it and prepare this exposition. Anthropic's public models were used in this process.

The authors did not receive financial compensation from Anthropic or PSI. This article is not intended for publication at a conference or in a journal. 
\end{abstract}

\newpage
\tableofcontents
\newpage
\pagenumbering{arabic}

\section{Introduction}

Quantum computing research seeks to discover and delineate the power of quantum computers over classical ones. Simon's and Shor's algorithms~\cite{Sim97,Sho97} provided early demonstrations of quantum advantage. On the other hand, despite decades of effort, $\NP$-complete problems and the lattice problems that underlie post-quantum cryptography appear resistant to efficient quantum and classical algorithms alike.

An oft-repeated mantra is that quantum speedups exploit {\sl structure} in the inputs. This intuition is sharpest in the query model, where many quantum algorithms operate. The crux of Shor's factoring algorithm is an algorithm for {\sc Period-Finding}, where one is given black-box query access to a string of integers that is promised to be periodic and is asked to find the period. Likewise, in Simon's problem one is given black-box query access to $f : \zo^n \to \zo^n$ that is promised to be two-to-one, with $f(x) = f(x\oplus s)$ for some nonzero $s$, and is asked to find $s$. In contrast, Grover's algorithm~\cite{Gro96} for searching an unordered list of $n$ items does so with $O(\sqrt{n})$ queries, achieving only a quadratic speedup over classical algorithms, a speedup that Bennett, Bernstein, Brassard, and Vazirani~\cite{BBBV97} showed is optimal.

\paragraph{The simulation conjecture.} Common to Shor's and Simon's algorithms, and many superpolynomial speedups that followed, is that they rely on a stringent promise on the input---one that is only satisfied by a tiny fraction of all inputs. The following conjecture formalizes a sense in which such a promise is necessary: 

\medskip 

 \begin{tcolorbox}[colback = white,arc=1mm, boxrule=0.25mm]

\begin{conjecture}[Simulation conjecture]\label{conj:simulation}
Let $\calA$ be a quantum algorithm that makes $t$ queries to an input $x\in\zo^n$. For every $\eps,\delta>0$ there is a classical algorithm that makes $\poly(t,1/\eps,1/\delta)$ queries to $x$ and approximates the acceptance probability of $\calA$ to within an additive $\eps$ on a $1-\delta$ fraction of inputs.
\end{conjecture}

\end{tcolorbox}
\medskip

This conjecture first appeared in the works of Aaronson~\cite{Aar05,Aar08talk} and of Aaronson and Ambainis~\cite{AA14}, where it is attributed to folklore.

The conjecture has an  equivalent formulation in the language of partial functions. Suppose $F : \zo^n \to \{0,1,*\}$ admits a $t$-query quantum algorithm. If $S \coloneqq F^{-1}(\zo)$ and $|S|\ge \delta 2^n$, the conjecture asserts the existence of a classical algorithm that makes $\poly(t,1/\delta,1/\alpha)$ queries and computes $F$ on a $1-\alpha$ fraction of $S$.   For total functions where $S = \zo^n$, an extreme case of entirely unstructured inputs, Beals, Buhrman, Cleve, Mosca, and de Wolf~\cite{BBCMW01} showed that their quantum and classical query complexities are indeed polynomially related.

\paragraph{The Aaronson--Ambainis (AA) conjecture.} Aaronson and Ambainis~\cite{AA14} reduced the simulation conjecture to a question about bounded low-degree polynomials: 

\medskip 

 \begin{tcolorbox}[colback = white,arc=1mm, boxrule=0.25mm]

\begin{conjecture}\label{conj:AA}
Let $p\colon\zo^n\to[0,1]$ be a polynomial of degree $d$. Then some variable $i\in[n]$ satisfies $\Inf_i(p)\ge\poly(\Var(p)/d)$.
\end{conjecture}

\end{tcolorbox}

\medskip

It is not hard to see why the simulation conjecture would follow from such a conjecture. Beals et al.~\cite{BBCMW01} showed that the acceptance probability of every $t$-query quantum algorithm is a polynomial $p$ of degree at most $2t$. A classical simulator can  query the most influential variable of $p$, restrict $p$ accordingly, and recurse on the restricted polynomial until it becomes close to constant. The lower bound on maximum influence asserted by the AA conjecture translates into a bound on the query complexity of this classical simulator. 

\paragraph{Implications for $\BPP$ vs.~$\BQP$ relative to a random oracle.} Further motivation for the AA conjecture comes from the relativized $\BPP$ versus $\BQP$ problem. Bernstein and Vazirani~\cite{BV97}, in the paper that defined $\BQP$, gave an oracle relative to which $\BPP\ne\BQP$. Fortnow and Rogers~\cite{FR99} asked whether the same holds relative to a {\sl random} oracle. This question remains open~\cite{Aar05,For21}.   

Aaronson and Ambainis showed that their conjecture, if true, would provide formal evidence for the difficulty of~\cite{FR99}'s question: Assuming the AA conjecture, proving  $\BQP\not\sse\mathsf{HeurP}$ relative to a random oracle would require proving the {\sl unrelativized} separation $\P \ne \P^{\# \P}$.\footnote{\cite{AA14} state their assumption as $\BQP \not\sse\AvgP$ relative to a random oracle, where by $\AvgP$ they mean the languages decided by a polynomial-time algorithm on a $1-o(1)$ fraction of the inputs. We state their result with $\mathsf{HeurP}$ instead of $\AvgP$ since the latter is often reserved for errorless algorithms.} This was recently strengthened by Blanc, Docter, Strassle, and Tan~\cite{BDST26}, who showed that it would require proving not only $\P \ne \P^{\# \P}$, but even $\PromiseBPP \ne \PromiseBQP$:\footnote{\cite{BDST26} state their assumption as $\PromiseBQP\not \sse\PromiseHeurBPP$ relative to a random oracle, which is implied by $\BQP \not \sse \mathsf{HeurBPP}$ relative to a random oracle. By a standard argument~\cite{BG81}, $\mathsf{HeurBPP}=\mathsf{HeurP}$  relative to a random oracle.}

\medskip 

 \begin{tcolorbox}[colback = white,arc=1mm, boxrule=0.25mm]

\begin{claim}
\label{claim:random oracle implication intro}
    Assume the AA conjecture. If $\BQP\not\sse\mathsf{HeurP}$ relative to a random oracle then  $\PromiseBPP \ne \PromiseBQP$.
\end{claim}
\end{tcolorbox}
\medskip 

\cite{BDST26}'s motivation for~\Cref{claim:random oracle implication intro} came from the fact that $\BPP \ne \BQP$ implies $\BPP \ne \BQP$ relative to a random oracle~\cite{BG81,FR99}. This ``easy direction," where an unrelativized separation implies a random oracle separation, holds for many pairs of complexity classes. \Cref{claim:random oracle implication intro} says that the AA conjecture implies an approximate  converse, one going in the ``hard direction," and hence an approximate equivalence between the status of the $\BPP$ vs.~$\BQP$ problem in the random oracle and unrelativized worlds (approximate because of the difference between $\mathsf{BPP}$ and $\mathsf{HeurP}$, and the distinction between promise and non-promise classes). In other words, the AA conjecture implies a sense in which the random oracle hypothesis~\cite{BG81}, while false in general~\cite{Kur82,Har85,CCGHHRR94}, does hold for the $\BPP$ vs.~$\BQP$ problem. 

\subsection{This article}

We give an exposition of a recent proof of the AA conjecture~\cite{Tur26}. The resolution of this conjecture implies the simulation conjecture and makes~\Cref{claim:random oracle implication intro} unconditional. We show:  

\medskip 

 \begin{tcolorbox}[colback = white,arc=1mm, boxrule=0.25mm]
\begin{theorem}[AA is true]\label{thm:intro-AA}
Let $p\colon\zo^n\to[0,1]$ be a polynomial of degree $d$. Then
\[
  \max_{i\in[n]}\Inf_i(p) \;\geq\; \Omega\left(\frac{\Var(p)^{8}}{d^{12}}\right).
\]
\end{theorem}
\end{tcolorbox}
\medskip 

The proof in~\cite{Tur26} achieves stronger parameters: $\max_i \Inf_i(p) \ge \Omega(\Var(p)^{5/2}/d\hspace{0.08em}^6)$. All the key ideas in our presentation are from the original proof. We strive to present the proof as simply as possible, to highlight its main conceptual steps, and to place each step in the context of the broader literature on the analysis of boolean functions and probabilistic combinatorics. As a result, our technical overview is fairly long, but the proof that we present is short.

\subsection{Prior work}

\paragraph{The Aaronson--Ambainis conjecture.} Dinur, Friedgut, Kindler, and O'Donnell~\cite{DFKO07}, in a paper that predates~\cite{AA14}, showed that every bounded degree-$d$ polynomial is $\eps$-close to a junta on $2^{O(d)}/\eps^2$ variables, which gives a variable of influence $\poly(\Var(p)/2^{d})$. This had remained the best bound for general polynomials, and has been rederived by O'Donnell and Zhao~\cite{OZ16} and by Defant, Masty\l{}o, and P\'erez~\cite{DMP19}. As~\cite{AA14} also noted in their paper, for {\sl boolean-valued} polynomials, the conjecture follows as a corollary of the ``OSSS inequality" of O'Donnell, Saks, Schramm, and Servedio~\cite{OSSS05} and bounds on decision tree depth in terms of polynomial degree~\cite{Nis91,NS94}. We  discuss the OSSS inequality in more detail in the technical overview.

 The conjecture has since been proven for several special cases: polynomials whose nonzero Fourier coefficients have the same magnitude~\cite{Mon12,Iva26,STV26}; completely bounded block-multilinear forms~\cite{BSdW22}, a class which contains the acceptance polynomials of $k$-fold {\sl Forrelation}~\cite{ABP19}; homogeneous Fourier completely bounded polynomials~\cite{EG24}; and random restrictions of arbitrary bounded polynomials~\cite{Bha25}.  O'Donnell and Zhao~\cite{OZ16} showed that it suffices to prove the conjecture for one-block decoupled polynomials, while Girish, Mittal, Nehoran, and Raz~\cite{GMNR26} showed that variants of the conjecture for certain non-uniform distributions are equivalent to the original one.

 Most relevant for us, Lovett and Zhang~\cite{LZ23} connected the problem to fractional certificates and proved the first bound on the fractional certificate complexity of bounded low-degree functions.  We elaborate on the connections between the proof that we present and~\cite{LZ23} in the technical overview.

\paragraph{The simulation conjecture.} The simulation conjecture concerns decision problems, and its natural analogue for {\sl search} problems is false. Yamakawa and Zhandry~\cite{YZ24} gave a search problem for which, relative to a random oracle $\mathcal{O} : \zo^n \to \zo^n$, there is a quantum algorithm that uses polynomially many queries and succeeds with high probability, whereas every classical algorithm requires exponentially many queries.

Recent works study the simulation conjecture for quantum algorithms with limited rounds of adaptivity~\cite{BDST26,EPS26,LM26}.

\section{Technical Overview}
\label{sec:technical overview}

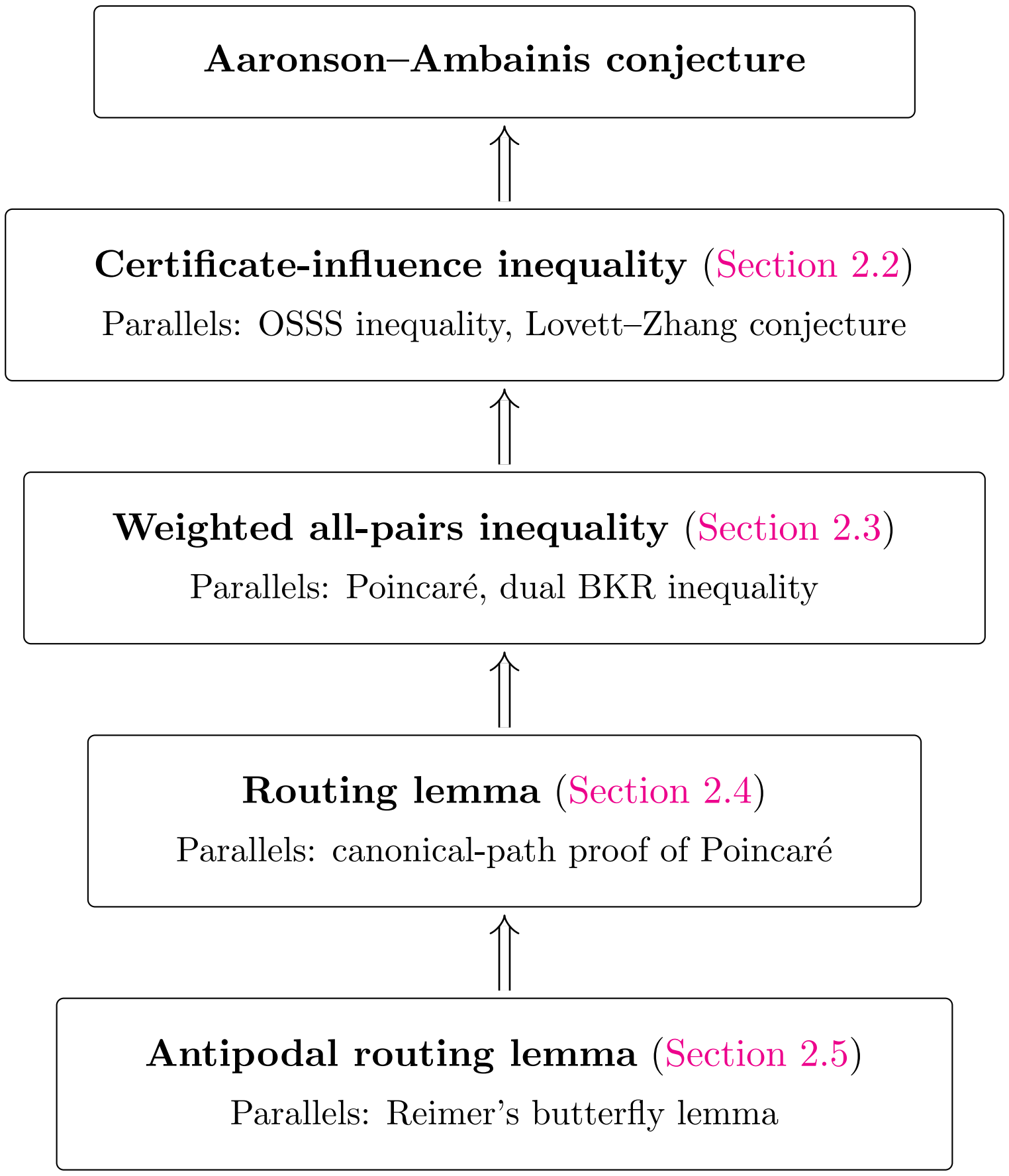
\begin{figure}[ht]
\centering
\begin{tikzpicture}[
  box/.style={draw, rounded corners=2pt, align=center, minimum width=8.2cm, inner xsep=25pt, inner ysep=11pt},
  imp/.style={-{Implies}, double, double distance=2.5pt, line width=0.5pt, shorten <=2pt, shorten >=2pt},
  node distance=9mm]
  \node[box] (aa) {\textbf{Aaronson--Ambainis conjecture}};
  \node[box, below=of aa] (ci) {\textbf{Certificate-influence inequality} (\Cref{sec:key-lemma}) \\[3pt] \small Parallels: OSSS inequality, Lovett--Zhang conjecture};
  \node[box, below=of ci] (wt) {\textbf{Weighted all-pairs inequality} (\Cref{sec:all pairs overview}) \\[3pt] \small Parallels: Poincar\'e, dual BKR inequality};
  \node[box,  below=of wt] (rt) {\textbf{Routing lemma} (\Cref{sec:routing overview})\\[3pt]  \small Parallels: canonical-path proof of Poincar\'e};
  \node[box, below=of rt] (mt) {\textbf{Antipodal routing lemma} (\Cref{sec:matching overview})\\[3pt] \small Parallels: Reimer's butterfly lemma};
  \draw[imp] (ci) -- (aa);
  \draw[imp] (wt) -- (ci);
  \draw[imp] (rt) -- (wt);
  \draw[imp] (mt) -- (rt);
\end{tikzpicture}
\caption{Structure of the proof.}
\label{fig:roadmap}
\end{figure}

\paragraph{Roadmap.} \Cref{fig:roadmap} summarizes the structure of the proof and of this technical overview. We will present the proof in a top-down fashion, devoting a subsection to each step. As the figure also indicates, each step draws on and extends ideas in the literature, which we discuss in the corresponding subsections. This section therefore doubles as an extended survey of related work. For a more succinct technical overview, a reader can skip all subsections titled ``Discussion:\,[...]." 
\medskip

\paragraph{Preliminaries.}  For a real number $\xi$ we write $\xi_+ = \max(\xi,0)$. We write $\Ind[\cdot]$ for the indicator of a condition. Random variables are written in boldface (e.g.~$\bx\sim \zo^n$), and unless otherwise stated, all probabilities and expectations are with respect to the uniform distribution.  For a vector $\lambda \in \R_{\geq 0}^n$ we write $\lambda(i)$ for its $i$-th entry, $\lambda(S) = \sum_{i\in S} \lambda(i)$ for  $S\sse[n]$, and $\Lone{\lambda} = \lambda([n])$. 

For $x\in \zo^n$ and $S \sse [n]$, we write $x^{\oplus S}$ to denote $x$ with the coordinates in $S$ flipped. For $i\in[n]$ we write $x^{\oplus i}$ for $x^{\oplus\{i\}}$. We write $\Delta(x,y) = \{ i \in [n] : x_i \neq y_i\}$ for the set of coordinates where $x$ and $y$ differ. For $x,y\in\zo^n$, we write
\[
  \cube(x,y) \coloneqq \big\{z\in\zo^n : z_i = x_i = y_i \text{ for every } i\notin\Delta(x,y)\big\}
\]
to denote the subcube spanned by $x$ and $y$, i.e.~the subcube with $x$ and $y$ as antipodal points. 

We reserve $f$ for boolean-valued functions and use $p$ for real-valued ones. The variance of a function $p : \zo^n \to \R$ is $\Var(p) \coloneqq \E[(p(\bx)-\E[p(\bx)])^2]$. For boolean-valued functions $f : \zo^n\to\zo$ we have $\Var(f) = \Pr[f(\bx)=1]\Pr[f(\bx)=0]$. 
 
\begin{definition}[Influences]\label{def:influence}
Let $p\colon \zo^n \to \R$ and $i \in [n]$. The {\sl influence} of coordinate $i$ on $p$ is
\[
  \Inf_i(p) \coloneqq \Ex_{\bx}\Big[\big(p(\bx) - p(\bx^{\oplus i})\big)^2\Big].
\]
The total influence of $p$ is $\Inf(p) \coloneqq \sum_{i=1}^n\Inf_i(p)$.
\end{definition}

\subsection{Fractional certificates} An important notion in this proof is that of fractional certificates. Fractional certificates for boolean-valued functions were first studied by Tal~\cite{Tal13} and Gilmer, Saks, and Srinivasan~\cite{GSS16}. Up to constant factors, they are equivalent to randomized certificates, introduced by Aaronson~\cite{Aar08}. 

We will need an analogue for real-valued functions: 

\begin{definition}[Fractional certificates]\label{def:certificate}
Let $p\colon \zo^n \to \R$ and $x \in \zo^n$. A {\sl fractional certificate for $p$ at $x$} is a vector $\lambda_x \in \R^n_{\geq 0}$ such that
\[
  \lambda_x\big(\Delta(x,y)\big) \ge |p(x) - p(y)|   \qquad \text{for every } y \in \zo^n.
\]\
Its {\sl cost} is $\Lone{\lambda_x}$.
For a boolean-valued function $f : \zo^n\to\zo$, this is equivalent to: 
\begin{equation*}
  \lambda_x\big(\Delta(x,y)\big) \geq 1 \qquad \text{for all $y$ such that } f(y) \neq f(x).
\end{equation*}
The $\zo$-valued vectors $\lambda_x$ satisfying this are exactly the indicator vectors of ordinary certificates of $f$ at $x$.
\end{definition}

\begin{definition}[Fractional certificate families]\label{def:certificate-family}
A {\sl fractional certificate family} for $p\colon\zo^n\to\R$ is an indexed collection $\Lambda = (\lambda_x)_{x\in\zo^n}$ in which $\lambda_x$ is a fractional certificate for $p$ at $x$, for every $x \in \zo^n$. Its {\sl cost} is $\max_x \Lone{\lambda_x}$.

The {\sl fractional certificate complexity} of $p$ at $x$, written $\FC(p,x)$, is the minimum cost of a fractional certificate for $p$ at $x$, and the {\sl fractional certificate complexity} of $p$ is $\FC(p) = \max_x\FC(p,x)$, the minimum cost of a fractional certificate family for $p$. 
\end{definition}

Agarwal and Ben-David~\cite{ABD26}, building on~\cite{ABK21,Tal13}, proved the following bound on the fractional certificate complexity of bounded low-degree polynomials: 

\begin{fact}\label{fact:ABD}
Every $p\colon\zo^n\to[0,1]$ of degree at most $d$ has $\FC(p) = O(d^2)$.
\end{fact}

Earlier, Lovett and Zhang~\cite{LZ23} proved that every such $p$ has $\FC_\eps(p) = O(d^8\log^{16}(n)/\eps^4)$. Here $\FC_\eps(p)$ is an analogue of $\FC(p)$ in which $x$'s certificate must satisfy $\lambda_x(\Delta(x,y))\geq1$ for every~$y$ with $|p(x)-p(y)|\geq\eps$, and is unconstrained for other $y$'s. A fractional certificate of cost $K$ yields one of cost $K/\eps$ in their sense, and so~\Cref{fact:ABD} implies $\FC_\eps(p)=O(d^2/\eps)$. We will use the stronger notion $\FC(p)$.

\subsection{Certificate-influence inequality}\label{sec:key-lemma}

\Cref{fact:ABD} is the only step in the  proof that uses the degree assumption of the AA conjecture. With this fact in hand, we show that every bounded function with small fractional certificate complexity, regardless of degree, has an influential variable. This claim in turn reduces to the following statement about boolean-valued functions:

\medskip 

 \begin{tcolorbox}[colback = white,arc=1mm, boxrule=0.25mm]

\begin{restatable}[Certificate-influence inequality]{lemma}{key}
\label{lem:key}
\label[certinf]{cert-inf-inequality}
Let $f\colon\zo^n\to\zo$ be nonconstant, with a fractional certificate family $\Lambda = (\lambda_x)_{x\in\zo^n}$. Then for uniform $\bx$ and $\by$,
\[
 \max_{i\in[n]} \Inf_i(f) \ge 4 \cdot \Ex_{\bx}\left[\frac{\Ind[f(\bx)=1]}{\Lone{\lambda_{\bx}}}\right]
  \cdot
  \Ex_{\by}\left[\frac{\Ind[f(\by)=0]}{\Lone{\lambda_{\by}}}\right]. 
\]
\end{restatable}
\end{tcolorbox}

\Crefformat{certinf}{#2certificate-influence inequality#3}

\medskip 
Note that the RHS is well defined: Since $f$ is nonconstant, every input requires a certificate of positive cost. The inequality is exactly tight for the dictator function $f(x) = \Ind[x_1 = 1]$ with $\lambda_x = e_1$ for every $x$.   To our knowledge, even the special case for ordinary (non-fractional) certificates was not known, though we need the stronger fractional version to prove the AA conjecture. 

\subsubsection{How the AA conjecture follows}
\label{sec:AA from cert-inf}

\begin{figure}[ht]
\centering
\colorlet{inA}{blue!60!black}
\colorlet{notA}{red!70!black}
\begin{tikzpicture}[x=9cm, y=1.6cm, lbl/.style={font=\small}]
  \def\dens(#1){0.3+0.6*exp(-((#1-0.15)/0.08)^2)+1.15*exp(-((#1-0.47)/0.085)^2)+0.42*exp(-((#1-0.78)/0.05)^2)+0.95*exp(-((#1-1.02)/0.1)^2)}
  \def\t{0.565}
  \begin{scope}
    \clip (0,0) rectangle (\t,2);
    \fill[notA!12] (0,0) -- plot[domain=0:1, samples=200] (\x, {\dens(\x)}) -- (1,0) -- cycle;
  \end{scope}
  \begin{scope}
    \clip (\t,0) rectangle (1,2);
    \fill[inA!12] (0,0) -- plot[domain=0:1, samples=200] (\x, {\dens(\x)}) -- (1,0) -- cycle;
  \end{scope}
  \def\w{0.07}
  \begin{scope}
    \clip (\t-\w,0) rectangle (\t+\w,2);
    \fill[pattern=north east lines, pattern color=green!50!black] (0,0) -- plot[domain=0:1, samples=200] (\x, {\dens(\x)}) -- (1,0) -- cycle;
  \end{scope}
  \draw[thick] plot[domain=0:1, samples=200] (\x, {\dens(\x)});
  \draw[semithick] (0,0) -- (1,0);
  \foreach \xx/\l in {0/0, 1/1}{
    \draw[semithick] (\xx,-0.07) -- (\xx,0.07);
    \node[below] at (\xx,-0.07) {$\l$};
  }
  \draw[semithick] (\t,-0.07) -- (\t,0.07);
  \node[below] at (\t,-0.07) {$t$};
  \node[lbl, anchor=west] (cert) at (0.66,1.13) {certificates};
  \node[lbl, anchor=south, inner sep=0pt] at ([yshift=-1.5pt]cert.north) {expensive new};
  \draw[-{Stealth[length=4pt]}, black!70] (cert.west) -- ({\t+0.014},{\dens(\t+0.014)+0.04});
  \draw[thick, gray, dash pattern=on \pgflinewidth off 1.25pt] (\t,0) -- (\t,2.05);
  \node[notA, anchor=south] at (\t/2,1.75) {$p_t(x) = 0$};
  \node[inA, anchor=south] at ({(1+\t)/2},1.75) {$p_t(x) = 1$};
\end{tikzpicture}
\caption{The curve depicts the distribution of $p(\bx)$ for a uniform $\bx\sim\zo^n$. We booleanize $p$ by thresholding it at a value $t\in [0,1]$. Since the certificates for $p_t$ are the certificates for $p$ scaled by $1/|p(x)-t|$, inputs with $p(x)$ close to $t$ (shaded green) have expensive new certificates. Such inputs contribute little to the RHS of the certificate-influence inequality.}
\label{fig:threshold}
\end{figure}
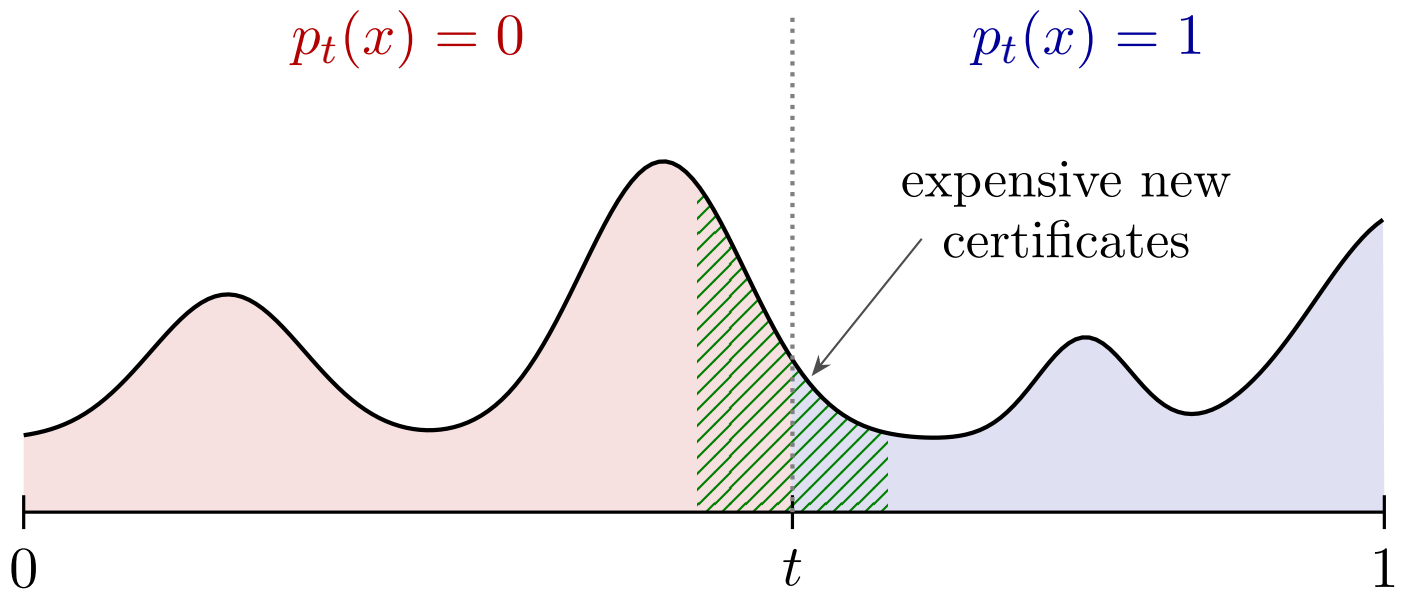

Let $p : \zo^n \to [0,1]$ be a bounded polynomial and $\Lambda = (\lambda_x)_{x\in \zo^n}$ be a fractional certificate family for $p$ of cost $\FC(p)$.  For $t \in [0,1]$, we define the boolean-valued threshold function 
\[ p_t(x) \coloneqq \Ind[p(x) > t].  \]
The high-level idea is that the certificate-influence inequality applied to $p_t$  yields an AA-like statement for $p_t$, which, for an appropriate choice of $t$, we can then transfer to $p$. See~\Cref{fig:threshold}.

It can be checked that 
\begin{equation} \left(\frac{\lambda_x}{|p(x)-t|}\right)_{x\in \zo^n}\label{eq:new certs}
\end{equation}
is a fractional certificate family for $p_t$. Applying the certificate-influence inequality to $p_t$ and these new certificates, we get that 
\begin{align*} \max_{i\in [n]}\Inf_i(p_t) 
&\ge 4 \cdot \frac{\E\big[ (p(\bx)-t)_{+}\big] \Ex\big[(t-p(\by))_{+}\big]}{\FC(p)^2}.
\end{align*}

\Cref{fact:ABD} upper bounds $\FC(p)$ by $O(\deg(p)^2)$, while $\E[ (p(\bx)-t)_{+}]$ and $\Ex[(t-p(\by))_{+}]$ are each at least $\Omega(\Var(p))$ as long as $t$ is sufficiently close to $\E[p(\bx)]$. Finally, for $\bt$ randomly chosen within this admissible interval, an elementary argument upper bounds $\E_{\bt}[\Inf_i(p_{\bt})]$ in terms of $\Inf_i(p)$ and $\Var(p)$. Putting these together yields the AA conjecture.

\subsubsection{Discussion: Average-case, median, and worst-case versions}
\label{sec:OSSS}
The certificate-influence inequality is an average-case bound: The RHS averages, over $1$-inputs and $0$-inputs separately, the reciprocals of the costs of their certificates. This implies a weaker  version where the lower bound is expressed in terms of the {\sl median} costs of $1$- and $0$-certificates: 

\begin{corollary}[Median version of the certificate-influence inequality]\label{cor:median}
Let $f$ and $\Lambda$ be as in \Cref{lem:key}. For $b\in\zo$, let $m_b$ be a {\sl median} of the cost $\Lone{\lambda_x}$ over the inputs $x$ with $f(x) = b$. Let $\mu \coloneqq \Pr[f(\bx)=1]$. Then
\[
\max_{i\in[n]}\Inf_i(f) \ge \frac{\mu(1-\mu)}{m_0 \cdot m_1}. 
\]
\end{corollary}

Our reduction from the AA conjecture can be modified so that this median version suffices, though we do not know of a proof of it that is substantially simpler than that of the average-case version.

\paragraph{Worst-case version and the OSSS inequality.} An even weaker version of the  inequality is one where the lower bound is expressed in terms of the {\sl maximum} costs of the certificates. This worst-case version, in contrast, does not suffice for our reduction from the AA conjecture. In our reduction, the inequality (or its median version) is applied to the new fractional certificates (\ref{eq:new certs}). The costs of these certificates blow up whenever $p(x)$ is close to $t$. An average-case or median version allows us to argue that this does not happen too often, but a worst-case version does not.

We also remark that, up to polynomial factors, such a worst-case version already follows from the OSSS inequality: 
\[ \max_{i\in [n]}\Inf_i(f) \ge \frac{4\mu(1-\mu)}{D(f)}\qquad \text{where $D(f) \coloneqq $ decision tree complexity of $f$}\] 
along with bounds on $D(f)$ in terms of $\FC(f)$ via~\cite{Nis91,BBCMW01}. The proof of~\Cref{lem:key} is rather different from, and quite a bit more involved than, the various known proofs of the OSSS inequality.

\subsubsection{Discussion: Comparison with a conjecture of Lovett and Zhang}
\label{sec:LZ}

Lovett and Zhang~\cite{LZ23} proposed a plan for the AA conjecture whose main step is a conjectured $L_\infty$ analogue of Talagrand's convex-distance inequality~\cite{Tal96}. We  now state and discuss their conjecture, describe a counterexample, and compare their conjecture to the certificate-influence inequality. 

\paragraph{\cite{LZ23}'s conjecture in the language of fractional certificates.} Roughly speaking, \cite{LZ23}'s conjecture says that if $X,Y\sse \zo^n$ are two disjoint sets such that every $x\in X$ is far in $L_\infty$ from $\mathrm{conv}(Y)$, then there must be a variable that is influential with respect to $X$ or to $Y$. See their paper for the precise formulation of their conjecture and the definition of this more general notion of influence.  Their reduction from the AA conjecture requires this level of generality, but we state here the special case of their conjecture where $X$ and $Y$ partition the cube: 

\begin{conjecture}[Special case of {\cite[Conjecture~37]{LZ23}}]\label{conj:LZ}
Let $f\colon\zo^n\to\zo$ be nonconstant, $A \coloneqq f^{-1}(1)$, and $\mu \coloneqq \Pr[f(\bx)=1] = |A|/2^n$. Then
\[
\max_{i\in[n]}\Inf_i(f) \ge \mu(1-\mu)\cdot   \poly\Big(\min_{x\in A}\; d_\infty\big(x,\mathrm{conv}(\Abar)\big)\Big).
\]
\end{conjecture}

The distance $d_\infty(x,\mathrm{conv}(\Abar))$ is exactly captured by fractional certificates: 

\begin{fact}
\label{fact:FC-dist}
  $d_\infty\big(x,\mathrm{conv}(\Abar)\big) = 1/\FC(f,x)$ for $x\in A$.  
\end{fact}
One direction of this identity ($\geq$) is proved in~\cite[Lemma~28]{LZ23}, while the other ($\le$) follows from LP duality.\footnote{\Cref{fact:FC-dist} gives a particularly nice reformulation of the certificate-influence inequality. It can be stated as
\begin{equation*}
    \max_{i\in[n]} \Inf_i(f) \ge 4 \cdot \Ex_{\bx}\left[d_{\infty}(\bx, \mathrm{conv}(f^{-1}(0)))\right] \cdot  \Ex_{\by}\left[d_{\infty}(\by, \mathrm{conv}(f^{-1}(1)))\right]. 
\end{equation*}}
Thanks to this identity,~\Cref{conj:LZ} can be equivalently stated as follows: 

\begin{conjecture}[\Cref{conj:LZ} in the language of fractional certificates]\label{conj:LZ for certs}
    Let $f:\zo^n\to\zo$ be nonconstant and $\mu \coloneqq \Pr[f(\bx)=1]$. Then 
    \begin{equation*} \max_{i\in [n]}\Inf_i(f) \ge \frac{\mu(1-\mu)}{\poly(\FC_1(f))},\label{eq:LZ for certs}\end{equation*}
    where $\FC_b(f) \coloneqq \max\{ \FC(f,x)\colon x\in f^{-1}(b)\}.$
\end{conjecture}

Notably, the lower bound is one-sided in the sense that it only depends on the fractional certificate complexities of the $1$-inputs, not those of the $0$-inputs.

\paragraph{A counterexample.} \cite{LZ23} showed that their conjecture would imply the AA conjecture. Unfortunately, their conjecture is already false in the special case described above.  Consider the {\sc Tribes} function with tribes of width $w \approx \log n - \log\ln n$, for which $\mu \approx \frac1{2}$. Every $x\in \textsc{Tribes}^{-1}(1)$ has a fractional certificate (in fact, an ordinary certificate) of cost $O(w) = O(\log n)$, and so~\Cref{conj:LZ for certs} asserts a lower bound of $1/\polylog(n)$ on maximum influence. This is false, since every variable has influence $O(\log(n)/n)$.

\paragraph{How the certificate-influence inequality differs from~\Cref{conj:LZ for certs}.} The certificate-influence inequality is two-sided: It depends on the certificate costs of both the $1$-inputs and the $0$-inputs. This is what lets it evade the counterexample, since every $0$-input of \textsc{Tribes} requires fractional certificate complexity $\Omega(n/w) = \Omega(n/\log n)$.

Two-sidedness alone, however, is not the new ingredient. A two-sided worst-case statement, \[ \max_{i\in [n]}\Inf_i(f)\geq\frac{\mu(1-\mu)}{\poly(\FC_0(f),\FC_1(f))},\] 
already follows from the OSSS inequality as discussed in~\Cref{sec:OSSS}. What is new is the average-case nature of the certificate-influence inequality, which, as discussed in the same section, is what our reduction from the AA conjecture needs. 

These differences aside, our reduction from the AA conjecture is similar in spirit to~\cite{LZ23}'s. For a polynomial $p: \zo^n \to [0,1]$, they apply their conjecture to the sets $X = \{ x \colon p(x) \ge t + \theta\}$ and $Y = \{ y \colon p(y) \le t- \theta\}$, while we apply the certificate-influence inequality to the  function $p_t(x) = \Ind[p(x) > t]$. In both cases, the AA conjecture for a real-valued function is reduced to an AA-like statement for a boolean function (a partial boolean function in their case).

\subsection{Weighted all-pairs inequality} 
\label{sec:all pairs overview}

\paragraph{Overlap.}  We call a pair $(x,y)$ of inputs {\sl bichromatic} if $f(x)\ne f(y)$. We will be interested in understanding the amount of ``overlap" between certificates of bichromatic pairs: 
\newcommand{\overlap}{\mathrm{overlap}}
\[ \overlap_{\Lambda}(x,y)\coloneqq \sum_{i\in \Delta(x,y)} \lambda_x(i)\lambda_y(i),\]
which is the inner product of $\lambda_x$ and $\lambda_y$ restricted to the coordinates on which $x$ and $y$ differ.

The ordinary (non-fractional) certificates of bichromatic pairs have overlap at least $1$: Viewed as sets, these certificates must intersect within $\Delta(x,y)$, since otherwise there will be an input that is consistent with both certificates. For fractional certificates, however, overlap can be much smaller. Consider the $n$-bit {\sc Majority} function, $x = 1^n$ and $y = 0^n$, and fractional certificates: 
\[ \lambda_x = \lambda_y = \big(\lfrac{2}{n},\ldots, \lfrac{2}{n}\big).\]
The overlap between these two certificates is only $O(\frac1{n})$. 

To prove the certificate-influence inequality, we will need a more general definition where overlap is measured with respect to positive weights $w\in \R^{n}_{> 0}$: 
\[ \overlap_{w,\Lambda}(x,y)\coloneqq \sum_{i\in \Delta(x,y)} w_i\lambda_x(i)\lambda_y(i). \]

We can now state the weighted all-pairs inequality:

\medskip 

 \begin{tcolorbox}[colback = white,arc=1mm, boxrule=0.25mm]

\begin{restatable}[Weighted all-pairs inequality]{lemma}{weightedinequality}
\label{lem:weighted}
Let $f\colon\zo^n\to\zo$ be nonconstant, with a fractional certificate family $\Lambda = (\lambda_x)_{x\in\zo^n}$. Let $w\in \R^n_{> 0}$. For independent and uniform $\bx$ and $\by$, 
    \[ \sum_{i=1}^n \frac{\Inf_i(f)}{w_i} \ge 4\cdot \Ex_{\bx,\by}\left[\frac{\Ind[f(\bx)=1, f(\by)=0]}{\overlap_{w,\Lambda}(\bx,\by)}\right]. \]
\end{restatable}
\end{tcolorbox}
\medskip

The ratio on the RHS is taken to be $0$ when the numerator is $0$. When the numerator is $1$ (i.e.~when $(x,y)$ is bichromatic), it can be checked that the overlap must be positive.\footnote{In fact, fractional certificates for bichromatic pairs satisfy $\sum_{i\in \Delta(x,y)}\min\{\lambda_x(i),\lambda_y(i)\}\ge 1$. See~\cite[Lemma 6.2]{KT16}.}

The certificate-influence inequality follows quite straightforwardly from the weighted all-pairs inequality, via an appropriate choice of weights and an application of Cauchy--Schwarz. (This is why we need the weighted version.) 

\subsubsection{Discussion: Relationship to the Poincar\'e and dual BKR inequalities}
\label{sec:poincare dual BKR}

Consider the {\sl unweighted} version of the all-pairs inequality where $w=1^n$, which already captures essentially all the complexity of the proof. It says that: 
\begin{equation*} \Inf(f) \ge 4\cdot \Ex_{\bx,\by}\left[\frac{\Ind[f(\bx)=1, f(\by)=0]}{\overlap_{\Lambda}(\bx,\by)}\right]. \label{eq:unweighted all pairs}
\end{equation*}

\paragraph{Poincar\'e inequality.} Unlike the certificate-influence inequality, the unweighted all-pairs inequality is known and elementary in the case of {\sl ordinary} certificates. Since the overlap between ordinary certificates of bichromatic pairs is at least $1$, the inequality follows from the classic Poincar\'e inequality over the hypercube: 
\[ \textbf{Poincar\'e:} \quad \Inf(f) \ge 4\mu(1-\mu), \quad \mu \coloneqq\Pr[f(\bx)=1].\] 
The challenge in proving the all-pairs inequality, even the unweighted version, stems from the fact that the overlap between {\sl fractional} certificates can be much smaller than $1$. 

The overlap can also be larger than~$1$, and so the all-pairs inequality is incomparable to  Poincar\'e. That said, en route to the all-pairs inequality we will prove a generalization of Poincar\'e.  See~\Cref{sec:routing overview}.

\paragraph{Dual BKR inequality.} 
One can consider a more general notion of ordinary certificates for two sets $X,Y \sse \zo^n$, extending the standard setting where $X$ and $Y$ partition the cube. For each $x\in X$, let $C_x\sse [n]$ certify $x$'s membership in $X$: all $z$'s that agree with $x$ on $C_x$ also belong to~$X$. Likewise, we assign to each $y\in Y$ a certificate $D_y\sse [n]$ and let $\Lambda$ be the family of all certificates. 

If $X$ and $Y$ are disjoint, then the overlap between the certificates for any  pair $(x,y)\in X\times Y$ must be at least $1$ (i.e.~their corresponding subcubes must be disjoint), for the same reason as in the standard setting. Now consider the setting where $X$ and $Y$ intersect, in which case these certificates can have $0$ overlap (i.e.~their corresponding subcubes may intersect). The  dual van den Berg--Kesten--Reimer (BKR) inequality of Kahn, Saks, and Smyth~\cite{KSS11} bounds the probability of this event:  
\[
  \Prx_{\bx,\by}\big[\bx\in X,\ \by\in Y,\ \overlap_{\Lambda}(\bx,\by) = 0\big] \;\leq\; \Pr[\bz \in X\cap Y].
\]
(Their actual inequality is stronger: A  pair $(x,y)$ counts on the LHS as long as there exist certificates for them that have $0$ overlap. That is, one does not have to fix the certificate family beforehand.) \cite{KSS11} used the dual BKR inequality to prove Rudich's conjecture~\cite{Rud88}, and Smyth~\cite{Smy01,Smy02} used it to resolve Tardos's conjecture~\cite{Tar89}. We refer the reader to these papers for details, mentioning here that both Rudich's and Tardos's conjectures are motivated by questions concerning random oracles, in the same way that the AA conjecture is connected to the question of whether $\BPP \ne \BQP$ relative to a random oracle via~\Cref{claim:random oracle implication intro}. 

While the statement of the all-pairs inequality is quite different from that of the dual BKR inequality, their proofs are related. 
The key ingredient in the proof of the dual BKR inequality is Reimer's butterfly lemma~\cite{Rei00}. The proof of our antipodal routing lemma yields a generalization of Reimer's butterfly lemma. See~\Cref{sec:reimer}.

\subsection{Routing lemma}
\label{sec:routing overview}

For the rest of this overview, it will be convenient to view $f : \zo^n \to \zo$ as the set $A\coloneqq f^{-1}(1)$. A pair $(x,y)$ is therefore bichromatic if exactly one of them lies in $A$. We write $\Inf_i(A)$ for $\Inf_i(f)$, and likewise $\Inf(A)$ for $\Inf(f)$.

\begin{definition}[Monochromatic faces] 
Fix $A\sse \zo^n$. For $x\in A$ and $T\sse[n]$, we call $T$ a {\sl monochromatic face of $x$} if $\cube(x,x^{\oplus T})$ is monochromatic. Equivalently, $x^{\oplus S}\in A$ for every $S\sse T$.  \end{definition}

We are now ready to state the routing lemma: 

\medskip 

 \begin{tcolorbox}[colback = white,arc=1mm, boxrule=0.25mm]
\begin{lemma}[Routing lemma]\label{lem:routing intro}
Let $A\sse\zo^n$. For every bichromatic pair $(x,y) \in A \times \Abar$, let $T = T(x,y)\sse\Delta(x,y)$ be a monochromatic face of $x$. Then one can choose, for every such pair $(x,y)$ and monochromatic face $T$, a point $z = z(x,y,T)\in A$ such that
\begin{enumerate}[label=$(\alph*)$]
  \item\label{item:between} $z\in\cube(x^{\oplus T},\,y)$.
  \item\label{item:close} For every $i \in [n]$, when $\bx,\by$ are uniform and independent on $\zo^n$,
  \[
    \Prx_{\bx,\by}\big[\bx\in A,\ \by\notin A,\ \bz_i\neq\by_i\big] \;\leq\; \lfrac14\cdot\Inf_i(A).
  \]
\end{enumerate}
\end{lemma}

\end{tcolorbox}
\medskip 

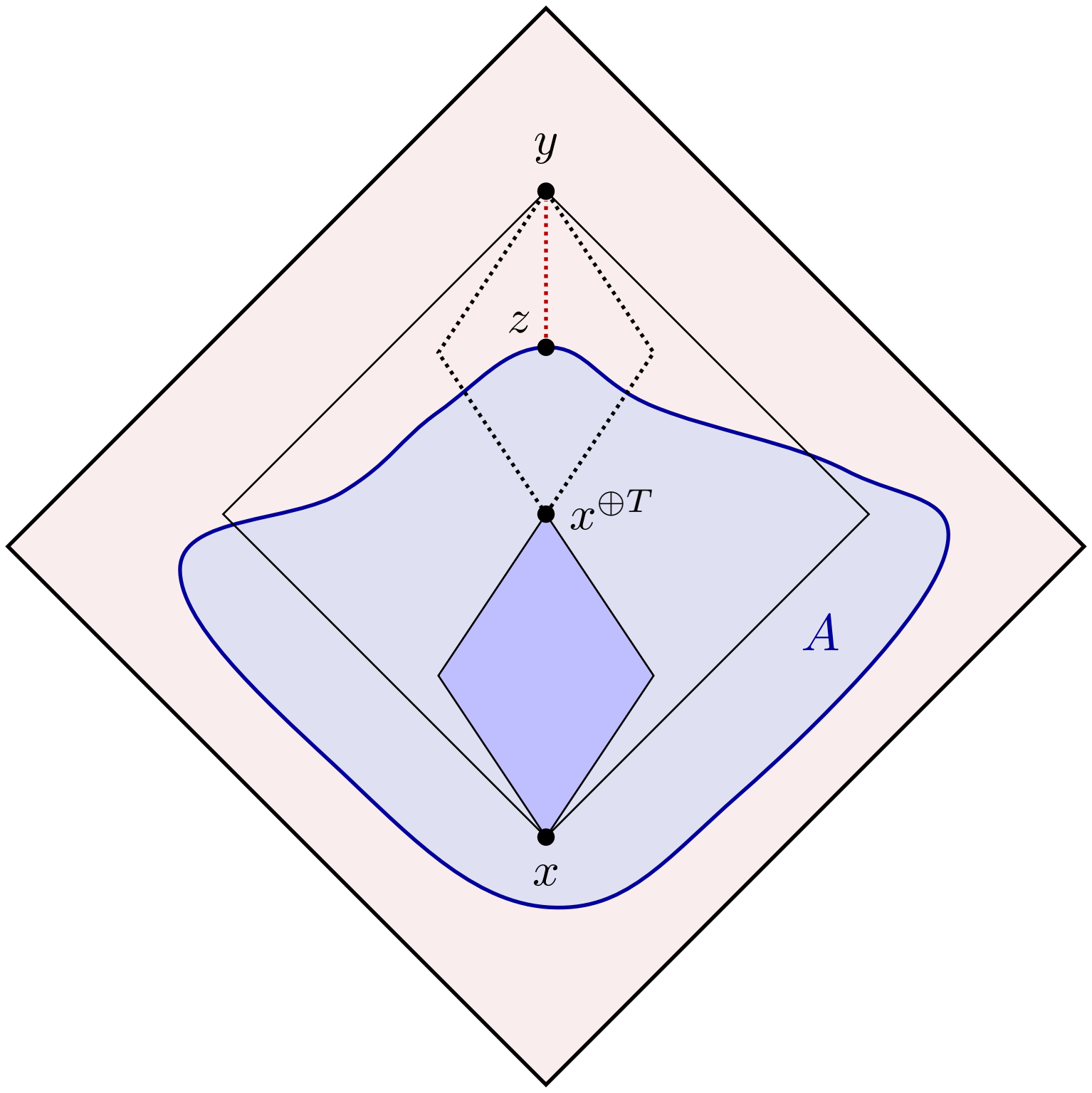
\begin{figure}[ht]
\centering
\scalebox{1.25}{%
\begin{tikzpicture}[scale=0.8,
  pt/.style={circle, fill, inner sep=1.3pt},
  lbl/.style={font=\small}]
  \fill[red!70!black!7] (0,-5.3) -- (5,-0.3) -- (0,4.7) -- (-5,-0.3) -- cycle;
  \draw[thick] (0,-5.3) -- (5,-0.3) -- (0,4.7) -- (-5,-0.3) -- cycle;
  \fill[blue!60!black!12] plot[smooth cycle, tension=0.7] coordinates
    {(-3.4,-0.5) (-1.9,0.2) (-1.0,0.95) (0,1.55) (1.0,1.0) (2.8,0.4) (3.7,-0.4) (1.8,-2.6) (0,-3.65) (-2,-2.3)};
  \draw[blue!60!black, thick] plot[smooth cycle, tension=0.7] coordinates
    {(-3.4,-0.5) (-1.9,0.2) (-1.0,0.95) (0,1.55) (1.0,1.0) (2.8,0.4) (3.7,-0.4) (1.8,-2.6) (0,-3.65) (-2,-2.3)};
  \node[blue!60!black] at (2.55,-1.1) {$A$};
  \draw (0,-3) -- (3,0) -- (0,3) -- (-3,0) -- cycle;
  \draw[fill=blue!25] (0,-3) -- (1,-1.5) -- (0,0) -- (-1,-1.5) -- cycle;
  \draw[densely dotted, thick] (0,0) -- (1,1.5) -- (0,3) -- (-1,1.5) -- cycle;
  \node[pt, label={[lbl]below:$x$}] at (0,-3) {};
  \node[pt, label={[lbl]above:$y$}] (y) at (0,3) {};
  \node[pt, label={[lbl, inner sep=1pt, xshift=1.7pt, yshift=1pt]right:$x^{\oplus T}$}] at (0,0) {};
  \node[pt, label={[lbl, inner sep=1.5pt]above left:$z$}] (z) at (0,1.55) {};
  \draw[red!70!black, densely dotted, thick] (z) -- (y);
\end{tikzpicture}%
}
\caption{An illustration of the routing lemma. The points $x$ and $y$ form a bichromatic pair, and $T$ is a monochromatic face of $x$: the shaded diamond $\cube(x,x^{\oplus T})$ lies in $A$. The dotted diamond is $\cube(x^{\oplus T},y)$. The routing lemma says that we can assign to $(x,y)$ and $T$ a point $z\in A\cap\cube(x^{\oplus T},y)$ that is ``close" to $y$---close, on average over all such $(x,y)$ pairs, in the coordinate-wise sense of~\ref{item:close}.   }
\label{fig:routing}
\end{figure}

See~\Cref{fig:routing} for an illustration of the lemma.

\subsubsection{Discussion: Poincar\'e for distances}
To gain some intuition for \ref{item:close}, consider the statement that one gets by summing it over all $i\in [n]$: 
\begin{enumerate}[label=$(b')$, leftmargin=1.5cm]
    \item\label{item:close-summed} $\ds \Ex_{\bx,\by}\big[\Ind[\bx \in A, \by\notin A]\cdot \dist(\bz,\by)\big] \le \lfrac{1}{4}\cdot \Inf(A)$, 
\end{enumerate}
where $\dist(\cdot,\cdot)$ denotes Hamming distance.  This says the Hamming distance from $z$ to $y$, averaged over all bichromatic $(x,y)$ pairs, is upper bounded in terms of $\Inf(A)$. The original statement \ref{item:close} says that this closeness even holds in a more stringent coordinate-wise manner, where for each $i\in [n]$, the disagreement in the $i$-th coordinate is upper bounded in terms of $\Inf_i(A)$.%

The summed version \ref{item:close-summed} alone, without \ref{item:between}, is already of independent interest. It implies: 
\[ \textbf{Poincar\'e for distances:} \quad  \Ex_{\bx,\by}\big[\Ind[\bx \in A, \ \by\notin A]\cdot \dist(\by,A)\big] \le \lfrac{1}{4}\cdot \Inf(A). \] 

This strengthens Poincar\'e, which can be written as: 
\[ \textbf{Poincar\'e:} \quad \Prx_{\bx,\by}\big[\bx\in A,\ \by\notin A\big] \le \lfrac1{4}\cdot \Inf(A).\] 
While Poincar\'e bounds the fraction of bichromatic pairs in terms of total influence, Poincar\'e for distances says that the same upper bound holds even when we take their distances to $A$ into account.

\subsubsection{Routing lemma \texorpdfstring{$\Rightarrow$}{implies} Weighted all-pairs inequality} \label{sec:overview-routing-to-weighted}

\paragraph{Poincar\'e for restricted distances.}
Taking~\ref{item:between} into consideration allows us to further strengthen Poincar\'e for distances, by restricting which paths from $y$ to $A$ we are allowed to take. Since $T\sse\Delta(x,y)$, requiring $z\in \cube(x^{\oplus T},y)$ is equivalent to requiring $\Delta(z,y) \sse \Delta(x,y)\setminus T$. Thus, we let
\[ \dist_{\Delta(x,y)\setminus T}(y,A)\coloneqq\dist\big(y,\ A\cap \cube(x^{\oplus T}, y)\big) \]
be the length of the shortest path from $y$ to $A$ that walks only in directions in $\Delta(x,y)\setminus T$. This is at least $\dist(y,A)$. Combining \ref{item:between} with \ref{item:close-summed} yields
\[ \textbf{Poincar\'e for restricted distances:} \quad  \Ex_{\bx,\by}\big[\Ind[\bx \in A, \ \by\notin A]\cdot \dist_{\Delta(\bx,\by)\setminus T}(\by,A)\big] \le \lfrac{1}{4}\cdot \Inf(A). \]
Recall that $T$ has to be a monochromatic face of $x$, and we can ensure this by ensuring $\lambda_{x}(T)<1$. Therefore, as long as a set of coordinates has $\lambda_{x}$-mass less than $1$, we may freeze those coordinates at their values in $y$ and force the path to only walk in the other directions.

\paragraph{Routing lemma $\Rightarrow$ unweighted all-pairs inequality.} We now sketch how the routing lemma implies the unweighted all-pairs inequality; the weighted version is similar. For every bichromatic pair $(x,y)$, we will define a monochromatic face $T= T(x,y) \sse \Delta(x,y)$ of~$x$ such that
\begin{equation} \dist_{\Delta(x,y)\setminus T}(y,A)\ \ge\ \frac{1}{\overlap_\Lambda(x,y)}. \label{eq:restricted dist}
\end{equation}
The inequality then follows by averaging and applying Poincar\'e for restricted distances.

Our goal is to define $T$ to make the restricted distance on the LHS of~\Cref{eq:restricted dist} large. Consider an arbitrary $z\in A$. Since $(y,z)$ is bichromatic, $\lambda_y(\Delta(y,z)) \ge 1$. Thus, if every coordinate $i\in\Delta(y,z)$ is ``light," in the sense that $\lambda_y(i)\le\tau$ for some $\tau>0$, then 
\[ |\Delta(y,z)|\ge \frac1{\tau}, \]
since at least $1/\tau$ coordinates of weight at most $\tau$ are needed to reach total weight $1$. This suggests that we should force the path from $y$ to $A$ to only walk along light directions, which we can do by freezing the heavy ones. Indeed, by letting 
\[ T=\{i\in\Delta(x,y)\colon \lambda_y(i)>\tau\}, \]
we get
\[ \dist_{\Delta(x,y)\setminus T}(y,A)\ge \frac1\tau. \]
It remains to choose $\tau$. A smaller $\tau$ gives a better bound, but the routing lemma needs $T$ to be a monochromatic face of $x$. Every $i\in T$ has $\lambda_y(i)>\tau$, so by Markov's inequality for the function $\lambda_y$ with respect to the measure $\lambda_x$,
$$\lambda_x(T) \;\le\; \frac{1}{\tau}\sum_{i \in T} \lambda_x(i)\,\lambda_y(i) \;\le\; \frac{1}{\tau}\sum_{i \in \Delta(x,y)} \lambda_x(i)\,\lambda_y(i) \;=\; \frac{\overlap_\Lambda(x,y)}{\tau},$$
where the first inequality is strict unless $\lambda_x(T) = 0$. We therefore choose $\tau = \overlap_\Lambda(x,y)$  to guarantee $\lambda_x(T)<1$. With this choice, we get~\Cref{eq:restricted dist}.

\begin{figure}[ht]
\centering
\colorlet{inA}{blue!60!black}
\colorlet{notA}{red!70!black}
\begin{tikzpicture}[
  br/.style={decorate, decoration={brace, amplitude=5pt, raise=2pt}},
  brm/.style={decorate, decoration={brace, mirror, amplitude=5pt, raise=2pt}},
  lbl/.style={font=\small}]
  \foreach \yy/\name/\mem/\col in {4.4/y/\notin A/notA, 1.95/z/{\in A\cap\cube(x^{\oplus T},y)}/inA, 0/x/\in A/inA}{
    \draw[\col, semithick, fill=white] (0,\yy-0.3) rectangle (8.76,\yy+0.3);
    \node[\col, anchor=west] at (8.96,\yy) {$\mem$};
    \node[\col, anchor=east] at (-0.15,\yy) {$\name =$};
  }
  \foreach \yy/\a/\b/\col in {4.4/1/1/notA, 1.95/1/{}/inA, 0/0/0/inA}{
    \foreach \k in {0,...,7}{
      \node[\col] at (1.69+0.38*\k-1.33,\yy) {$\a$};
      \node[\col] at (5.07+0.38*\k-1.33,\yy) {$\b$};
    }
  }
  \foreach \k/\v in {0/1,1/0,2/0,3/0,4/0,5/0,6/0,7/1}{
    \node[inA] at (5.07+0.38*\k-1.33,1.95) {$\v$};
  }
  \draw[thick, gray, dash pattern=on \pgflinewidth off 1.25pt] (3.38,5.3) -- (3.38,-0.3);
  \draw[thick, gray, dash pattern=on \pgflinewidth off 1.25pt] (6.76,5.3) -- (6.76,-0.3);
  \draw[br] (0,4.78) -- node[lbl, above=7pt] {$T$} (3.33,4.78);
  \draw[br] (0,5.55) -- node[lbl, above=7pt] {$\Delta(x,y)$} (6.76,5.55);
  \draw[brm] (0,4.02) -- node[lbl, below=7pt, align=center] {$\lambda_y(i) > \tau$\\{\footnotesize (heavy coordinates)}} (3.33,4.02);
  \draw[brm] (3.43,4.02) -- node[lbl, below=7pt, align=center] {$\lambda_y(i) \le \tau$\\{\footnotesize (light coordinates)}} (6.76,4.02);
  \draw[brm] (4.02,1.58) -- node[lbl, below=7pt] {$|\Delta(y,z)|\ge \frac{1}{\tau}$} (6.12,1.58);
  \draw[brm] (0,-0.38) -- node[lbl, below=7pt, align=center] {$\lambda_x(T) < 1$\\{\footnotesize (monochromatic face)}} (3.33,-0.38);
\end{tikzpicture}
\caption{How the routing lemma implies the (unweighted) all-pairs inequality. The three strings agree outside $\Delta(x,y)$, and $x$ and $y$ are drawn as all-$0$ and all-$1$ on $\Delta(x,y)$ for concreteness. The heavy coordinates $T$ of $\lambda_y$ form a monochromatic face of $x$ if $\tau = \overlap_\Lambda(x,y)$, and $z$ is the point given by the routing lemma.}
\label{fig:routing implies all pairs}
\end{figure}
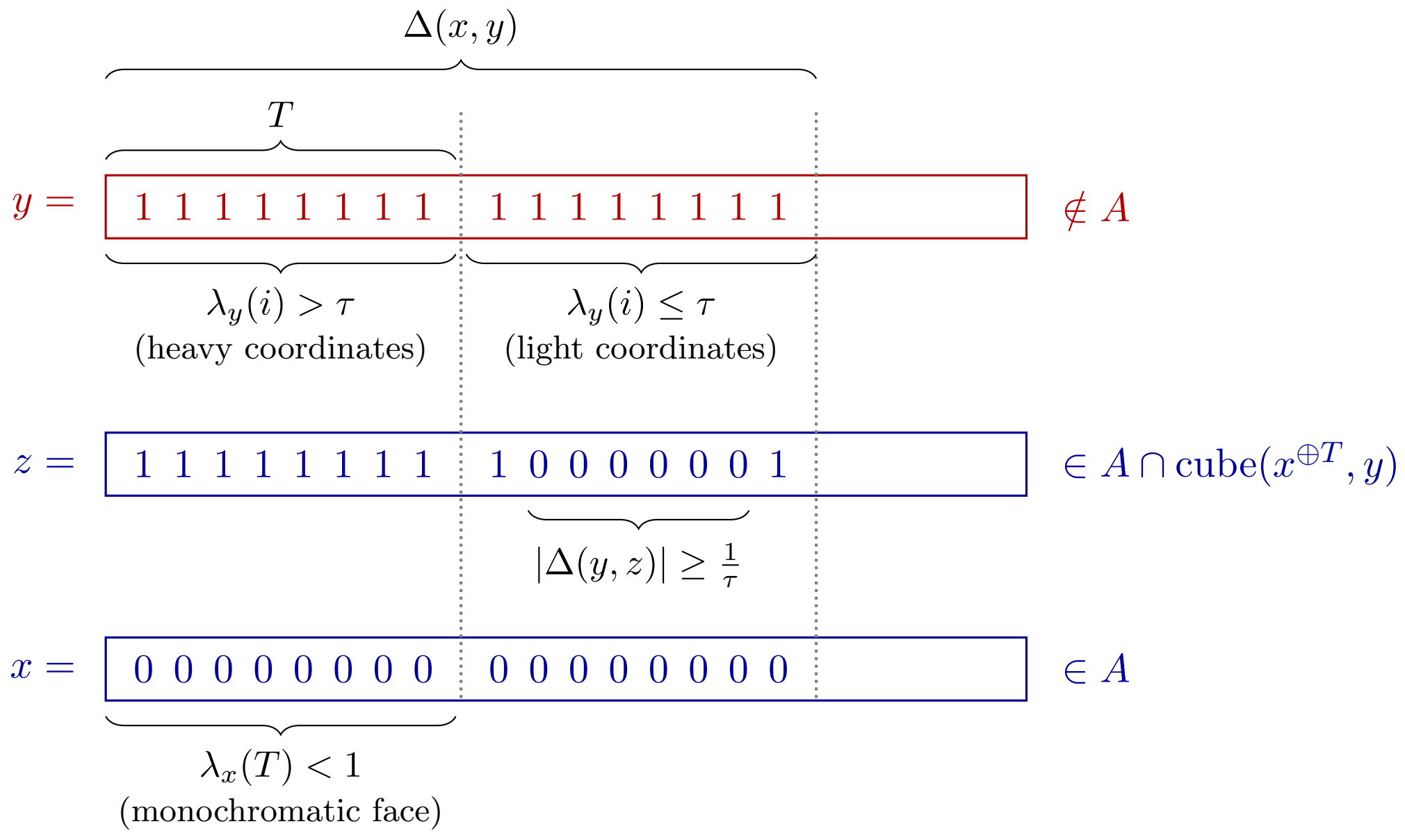

See~\Cref{fig:routing implies all pairs} for an illustration of this argument.

\subsection{Antipodal routing lemma}
\label{sec:matching overview}

 The crux of the proof of the routing lemma is the following analogue for {\sl antipodal} pairs $(x,\bar{x})$ where $\bar{x} \coloneqq (\bar{x}_1,\ldots,\bar{x}_n)$:

\medskip 

 \begin{tcolorbox}[colback = white,arc=1mm, boxrule=0.25mm]
\begin{lemma}[Antipodal routing lemma]\label{lem:matching intro}
Let $A\sse\zo^n$ and for every $x \in A$, let $T = T(x)\sse[n]$ be a monochromatic face of $x$. Then one can choose, for every such $x$ and monochromatic face $T$, a point $z = z(x,T)\in A$ such that
\begin{enumerate}[label=$(\alph*)$]
  \item\label{item:match-between} $z\in\cube(x^{\oplus T},\,\bar{x})$.
  \item\label{item:match-close}  For every $i\in[n]$, when $\bx$ is uniform on $\zo^n$,
  \[
    \Prx_{\bx}\big[\bx\in A,\ \bz_i\neq\bar{\bx}_i\big] \;\leq\; \lfrac12\cdot\Inf_i(A).
  \]
\end{enumerate}
\end{lemma}

\end{tcolorbox}

\medskip 

The routing lemma follows from this antipodal version by an averaging argument: Every bichromatic pair $(x,y)$ is an antipodal pair in $\cube(x,y)$, and we apply the antipodal routing lemma in every such subcube.  

\subsubsection{Discussion: Simple proof of the faceless antipodal routing lemma}

The main difficulty in proving the antipodal routing lemma, and hence the routing lemma, lies in the constraint imposed by the monochromatic faces: the requirement \ref{item:match-between} that $z\in \cube(x^{\oplus T}, \bar{x})$. We sketch a short and elementary proof for the special case where $T(x) = \varnothing$ for all $x$:

\begin{proposition}[Faceless antipodal routing lemma]
\label{prop:faceless matching}
    Let $A\sse\zo^n$. One can choose, for  each $x \in A$, a point $z=z(x)\in A$  such that 
  \[
 \Prx_{\bx}\big[\bx\in A,\ \bz_i\neq\bar{\bx}_i\big] \;\leq\; \lfrac12\cdot\Inf_i(A) \quad \text{for all $i\in [n].$}
  \]
\end{proposition}

\begin{proof}[Proof sketch]
    For $i\in[n]$, let $\pi_i\colon A\to A$ swap $x$ and $x^{\oplus i}$ (i.e.~flip their $i$-th coordinates) iff both lie in $A$, and let $\pi \coloneqq \pi_n\circ\cdots\circ\pi_1$. Our map is $x \mapsto \pi(x)$, which we furthermore claim is in fact a permutation of $A$.  A straightforward induction shows that for all $m\in [n]$, the map $\pi_{\le m} \coloneqq \pi_m\circ\cdots\circ\pi_1$  is indeed a permutation of $A$ and satisfies: 
\begin{equation*}
  \big|\{x\in A \colon \pi_{\le m}(x)_i \ne  \bar{x}_i\}\big| \;=\; \big|\{ x \in A \colon x^{\oplus i} \notin A\}\big|  \quad\text{for all $i\in[m]$.}
\end{equation*}
The $m=n$ case of the identity above, divided by $2^n$ on both sides, yields the proposition. 
\end{proof}

\paragraph{Relationship to the canonical-path proof of Poincar\'e.} Recall the following standard proof of the Poincar\'e inequality: For every pair $(x,y)$, walk from $x$ to $y$ by flipping the coordinates of $\Delta(x,y)$ in the fixed order $1 \to 2 \to \cdots \to n$. If the pair is bichromatic, this path must cross a bichromatic {\sl edge} of $A$, and we charge the pair to the first such edge. Each edge of the cube lies on these ``canonical paths" of exactly $2^{n}$ of the $4^n$ pairs, so the number of bichromatic pairs is at most $2^{n}$ times the number of bichromatic edges, which is the Poincar\'e inequality.

The proof of~\Cref{prop:faceless matching} also walks toward a target, the antipodal point $\bar{x}$, flipping coordinates in the fixed order $1 \to 2 \to \cdots \to n$, and charges bichromatic edges direction by direction. It differs in two ways. First, it never leaves $A$: It skips any flip that would exit $A$ (i.e.~it skips any bichromatic edge), and ends at the point $z = \pi(x)\in A$ rather than at $\bar{x}$. Second, every skipped flip is charged, not only the first: If the walk from $x$ is at $w$ and skips direction $i$, then $x$ is charged the bichromatic edge $\{w, w^{\oplus i}\}$. Thus $x$ incurs one charge for each coordinate on which $z$ and $\bar{x}$ differ. Since each step $\pi_i$ is a permutation of $A$, every bichromatic edge in direction $i$ is charged by exactly one $x$, which is where the equality in the proof comes from.

\paragraph{The actual antipodal routing lemma with faces.} The face constraint~\ref{item:match-between} requires $z$ to flip every coordinate $i\in T(x)$, which is not satisfied by this proof, and we know of no argument along these lines.%
Its proof, which we discuss next, is linear algebraic, and entirely different from any proof of the Poincar\'e inequality that we know.

\subsubsection{Two matrices and their determinants}
\label{sec:proof of matching lemma}

For a map $\pi : A \to A$, we let  
\[ \errors_i(\pi) \coloneqq \sum_{x \in A} \Ind[\pi(x)_i \neq \bar{x}_i] = 2^n \cdot \Prx_{\bx}\big[\bx\in A,\ \pi(\bx)_i\neq\bar{\bx}_i\big]. \] 
Consider the following natural approach. First construct a map $\pi$ satisfying \ref{item:match-close} (as we did in the proof of the faceless version), and then show that {\sl any} such map can be ``repaired" so that it furthermore satisfies the face constraints \ref{item:match-between} while not increasing $\errors_i(\pi)$ for any direction $i\in [n]$ (thereby preserving \ref{item:match-close}). Unfortunately, there are counterexamples showing that not all such maps can be repaired this way. One can then hope that any {\sl permutation} of $A$ can be repaired this way; after all, our proof of the faceless version constructs a permutation. But there are counterexamples to this as well.\footnote{Counterexamples for $n$ as small as $3$ can be easily found with your AI system of choice.}

Nevertheless, we show that a suitable modification of this high-level approach succeeds. The magical step is to take a {\sl signed combination of all possible permutations}. This makes some of them ``cancel" with each other. We show that {\sl any} permutation that survives this cancellation can indeed be repaired to satisfy the face constraints without increasing $\errors_i(\pi)$ for any direction $i\in [n].$

In more detail, we first define a matrix $M$ in formal variables $\zeta_1,\ldots,\zeta_n$ such that 
\[ \det M = \sum_{\text{permutations }\pi\text{ of }A} \sign(\pi) \prod_{i \in [n]} \zeta_i^{\errors_i(\pi)}. \]

We show that $\det M$ (viewed as a polynomial in the $\zeta_i$'s) has a ``good" monomial, one whose degrees, suitably normalized, are upper bounded in terms of the influences of $A$.  Here and throughout, when we refer to a monomial of a polynomial, we mean one with nonzero coefficient: 
\begin{claim}[There is a good monomial]
    \label{claim:nonzero-good}
    The polynomial $\det M$ has a monomial $\zeta_1^{k_1}\cdots \zeta_n^{k_n}$ such that
    \begin{equation}
        \label{eq:monomial-good}
        \frac{k_i}{2^n} \leq \lfrac12\cdot\Inf_i(A) \quad\quad\text{for every }i \in [n].
    \end{equation}
\end{claim}

Since a monomial $\zeta_1^{k_1}\cdots\zeta_n^{k_n}$ of $\det M$ has a nonzero coefficient, it arises from at least one permutation $\pi$ with $\errors_i(\pi)=k_i$ for every $i$, and $\errors_i(\pi)/2^n$ is exactly the LHS of~\ref{item:match-close} for $z=\pi$. So~\eqref{eq:monomial-good} tells us that this $\pi$ satisfies~\ref{item:match-close}.

Next, we define an analogous matrix $M_{\mcT}$ that takes the monochromatic faces $\mcT \coloneqq (T(x))_{x\in A}$ into account. Note that the face constraint $\pi(x) \in \cube(x^{\oplus T}, \bar{x})$ is equivalent to the requirement that $\pi(x)$ agrees with $\bar{x}$ on all coordinates in $T$: 
\begin{equation*}
    \det M_{\mcT} = \sum_{\text{permutations }\pi\text{ of }A} \underbrace{\Ind[\pi(x)_i = \bar{x}_i\text{ for all }x \in A, i \in T(x)]}_{\text{face constraint \ref{item:match-between}}}  \cdot \,\sign(\pi)  \prod_{i \in [n]} \zeta_i^{\errors_i(\pi)}.
\end{equation*}

We show that the existence of a good monomial in $\det M$ implies the existence of one in $\det M_{\mcT}$: 
\begin{claim}[Good monomial $\Rightarrow$ Good face-respecting monomial]
    \label{claim:nonzero-transfer}
    Let $\mcT = (T(x))_{x\in A}$ specify monochromatic faces for each $x\in A$. 
    For any monomial $\zeta_1^{k_1}\cdots \zeta_n^{k_n}$ of $\det M$ there is a monomial $\zeta_1^{k_1'}\cdots \zeta_n^{k_n'}$ of $\det M_{\mcT}$ such that
\[ 
        k_i' \leq k_i\quad\quad\text{for every }i \in [n].
   \]
\end{claim}
Applying \Cref{claim:nonzero-transfer} to the monomial given by \Cref{claim:nonzero-good} yields a monomial of $\det M_{\mcT}$, and hence a permutation satisfying \ref{item:match-between} and \ref{item:match-close}. This is the antipodal routing lemma.

\subsubsection{The determinant equivalence lemma}

The key lemma underlying both \Cref{claim:nonzero-good,claim:nonzero-transfer} is that $\det M$ and $\det M_{\mcT}$ have exactly the same {\sl minimal} monomials, defined as follows: 
\begin{definition}[Pareto-minimal component of a polynomial]
    \label{def:minimal}
    We say a monomial $\zeta_1^{k_1}\cdots \zeta_n^{k_n}$ is \textsl{Pareto-minimal} in a polynomial $p$ if there is no other monomial $\zeta_1^{k_1'}\cdots \zeta_n^{k_n'}$ of $p$ for which $k_i' \le k_i$ for all $i\in [n]$. We then denote the {\sl Pareto-minimal component} of $p$ as
    \begin{equation*}
        \minimal(p) \coloneqq \sum_{\text{monomials $m$ in $p$}}\Ind[\text{$m$ is Pareto-minimal in $p$}] \cdot c_m\, m,
    \end{equation*}
    where $c_m$ is $m$'s coefficient in $p$.
\end{definition}

\begin{restatable}[Determinant equivalence lemma]{lemma}{detequiv}
    \label{lem:det-equiv}    Let $\mcT = (T(x))_{x\in A}$ specify monochromatic faces for each $x\in A$. Then 
    \begin{equation*}
        \minimal(\det M) = \minimal(\det M_{\mcT}).
    \end{equation*}
\end{restatable}

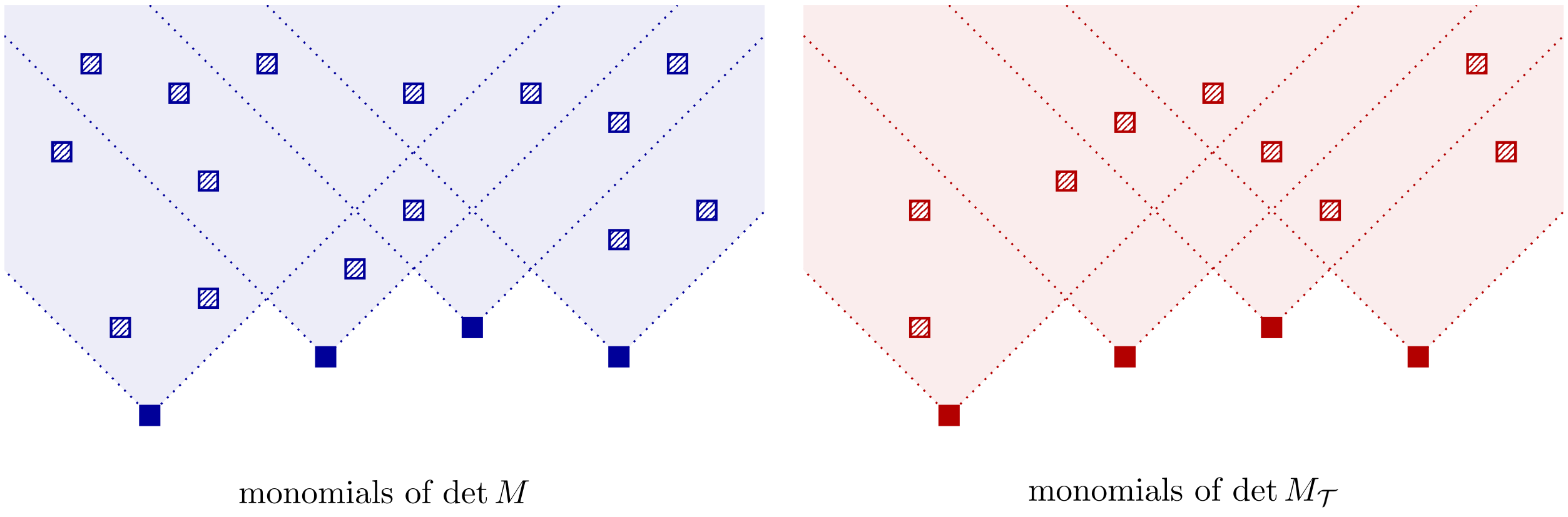
\begin{figure}[ht]
\centering
\begin{tikzpicture}[
  x=0.31cm, y=0.31cm,
  mono/.style={draw=\dmPc, thick, preaction={fill=white}, pattern={Lines[angle=45,distance=1.4pt,line width=0.45pt]}, pattern color=\dmPc, minimum size=5.6pt, inner sep=0pt, rectangle},
  pmin/.style={draw=\dmPc, thick, fill=\dmPc, minimum size=5.6pt, inner sep=0pt, rectangle},
  cone/.style={\dmPc, semithick, dotted},
  conefill/.style={fill=\dmPc!7}]
  \def\dmH{22}
  \def\dmPt#1#2{({(#2)-(#1)},{(#1)+(#2)})}
  \def\dmPanel#1#2#3{%
    \begin{scope}[xshift=#1]
      \begin{scope}
      \clip (-12.95,7) rectangle (12.95,\dmH);
      \foreach \a/\b in {8/0,6/4,4/7,1/9}{
        \pgfmathsetmacro{\X}{\b-\a}\pgfmathsetmacro{\Y}{\a+\b}
        \fill[conefill] (\X,\Y) -- (\X-\dmH+\Y,\dmH) -- (\X+\dmH-\Y,\dmH) -- cycle;}
      \foreach \a/\b in {8/0,6/4,4/7,1/9}{
        \pgfmathsetmacro{\X}{\b-\a}\pgfmathsetmacro{\Y}{\a+\b}
        \draw[cone] (\X-\dmH+\Y,\dmH) -- (\X,\Y) -- (\X+\dmH-\Y,\dmH);}
      \end{scope}
      \foreach \a/\b in {#2}{\node[mono] at \dmPt{\a}{\b} {};}
      \foreach \a/\b in {8/0,6/4,4/7,1/9}{\node[pmin] at \dmPt{\a}{\b} {};}
      \node[font=\small] at (0,5.4) {#3};
    \end{scope}}
  \def\dmPc{blue!60!black}\dmPanel{0cm}{7/8,5/15,5/13,7/12,15/5,13/6,10/1,12/8,11/5,14/3,9/10,9/3,7/6,3/11,2/13}{monomials of $\det M$}
  \def\dmPc{red!70!black}\dmPanel{8.45cm}{10/1,9/10,10/8,7/10,12/3,10/6,3/14,5/15,5/10}{monomials of $\det M_{\mcT}$}
\end{tikzpicture}
\caption{An illustration of the determinant equivalence lemma. The Pareto-minimal monomials of $\det M$, shaded solid blue, are the same as the Pareto-minimal monomials of $\det M_{\mcT}$, shaded solid red. Their other monomials may differ.}
\label{fig:det-equiv}
\end{figure}

See~\Cref{fig:det-equiv} for an illustration of the lemma. 

\Cref{claim:nonzero-transfer} follows directly from the determinant equivalence lemma. Indeed, it suffices to consider only Pareto-minimal monomials of $\det M$, and the determinant equivalence lemma says that every such monomial is also present in $\det M_{\mcT}$. 

We also use the determinant equivalence lemma to prove \Cref{claim:nonzero-good}. However, instead of applying it with the monochromatic faces specified by the antipodal routing lemma, we apply it with monochromatic faces that we define. We actually prove the stronger statement that \textsl{every} Pareto-minimal monomial of $\det M$ satisfies the degree bound of \Cref{eq:monomial-good}. The advantage of this stronger \textsl{for all} statement is that we can separately show, for each $i\in [n]$, that the degree bound is satisfied. For any $i$, we choose monochromatic faces $\mcT^{(i)}$ so that the $\zeta_i$-degree of every monomial in $\det M_{\mcT^{(i)}}$ satisfies the bound of~\Cref{eq:monomial-good}. By the determinant equivalence lemma, this must also be true of every Pareto-minimal monomial of $\det M$. Applying this argument $n$ times, once for each $i\in [n]$, yields~\Cref{claim:nonzero-good}.

\pparagraph{Proof of the determinant equivalence lemma.} The key step in the proof of \Cref{lem:det-equiv} is a linear-algebraic identity expressing $M_{\mcT}$ in terms of $M$. First, we define the two matrices, both with rows and columns indexed by elements of $A$:
\begin{equation*}
    M[x,z] \coloneqq \prod_{i \in [n]} \zeta_i^{\Ind[x_i = z_i]} \quad\quad \text{and}\quad\quad
    M_{\mcT}[x,z] \coloneqq \begin{cases}
        0 &\text{if }x_i = z_i\text{ for any }i \in T(x)\\
        M[x,z]&\text{otherwise.}
    \end{cases} 
\end{equation*}

Thus $M_{\mcT}$ is $M$ with certain entries zeroed out. Using the fact that each $T(x)\in \mcT$ is a monochromatic face of $x$, we show that there is in fact a closer relationship between the two matrices: Every row of $M_{\mcT}$, appropriately scaled, can be written as a linear combination of rows in $M$. 
\begin{equation}
    \label{eq:row-full}
    \paren[\Bigg]{\,\prod_{i \in T(x)} (1-\zeta_i^2)} \cdot \row_x(M_{\mcT}) = \sum_{S \subseteq T(x)} \paren[\Bigg]{\,\prod_{i \in S}(-\zeta_i)} \cdot \row_{x^{\oplus S}}(M).
\end{equation}
Note that the RHS is only well defined because $T(x)$ is a monochromatic face: The rows of $M$ are indexed by elements of $A$, so $x^{\oplus S}$ only appears as a row in $M$ if $x^{\oplus S}\in A$.  

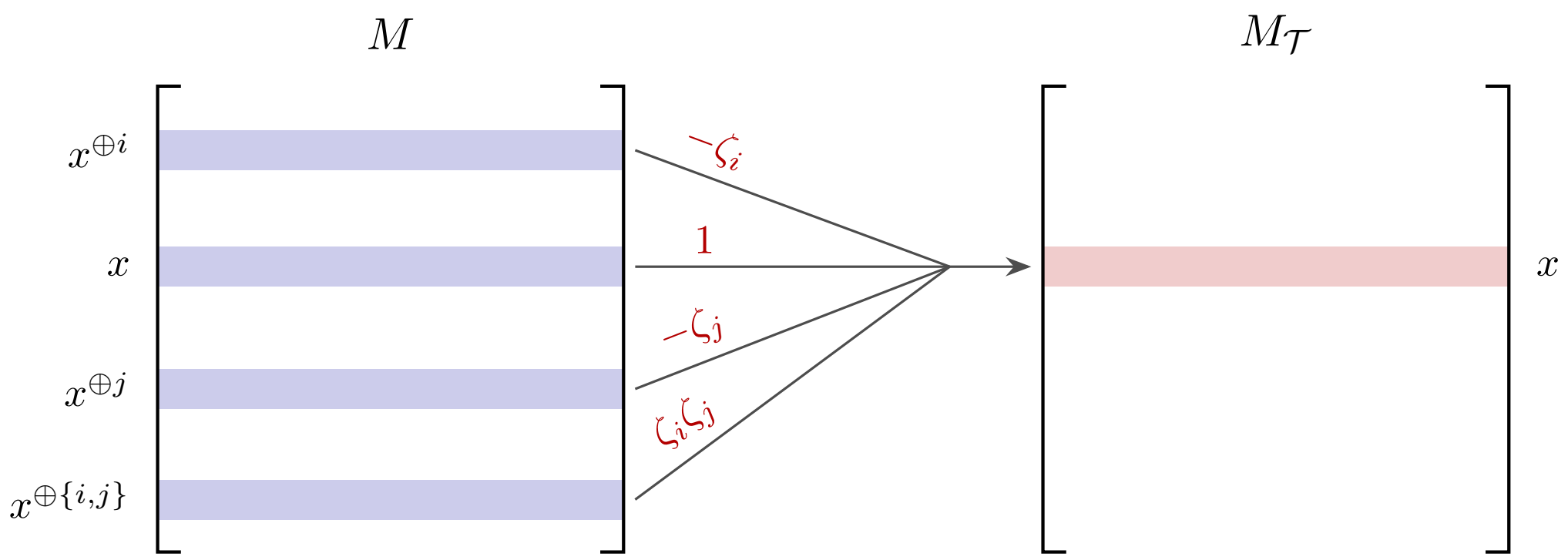
\begin{figure}[ht]
\centering
\colorlet{coefcol}{red!70!black}
\begin{tikzpicture}[lbl/.style={font=\small}]
  \def\hh{0.17}
  \foreach \yy in {3.45,2.45,1.4,0.45}
    \fill[blue!60!black!20] (0,\yy-\hh) rectangle (4,\yy+\hh);
  \draw[thick] (0.2,0) -- (0,0) -- (0,4) -- (0.2,4);
  \draw[thick] (3.8,0) -- (4,0) -- (4,4) -- (3.8,4);
  \node at (2,4.45) {$M$};
  \foreach \yy/\nm in {3.45/x^{\oplus i}, 2.45/x, 1.4/x^{\oplus j}, 0.45/{x^{\oplus \{i,j\}}}}
    \node[lbl, anchor=east] at (-0.1,\yy) {$\nm$};
  \fill[red!70!black!20] (7.6,2.45-\hh) rectangle (11.6,2.45+\hh);
  \draw[thick] (7.8,0) -- (7.6,0) -- (7.6,4) -- (7.8,4);
  \draw[thick] (11.4,0) -- (11.6,0) -- (11.6,4) -- (11.4,4);
  \node at (9.6,4.45) {$M_{\mcT}$};
  \node[lbl, anchor=west] at (11.7,2.45) {$x$};
  \foreach \yy/\cf in {3.45/-\zeta_i, 2.45/1, 1.4/-\zeta_j, 0.45/\zeta_i\zeta_j}
    \draw[semithick, gray!60!black] (4.1,\yy) -- node[pos=0.22, sloped, above, inner sep=3pt, lbl, text=coefcol] {$\cf$} (6.8,2.45);
  \draw[-{Stealth[length=2.2mm]}, semithick, gray!60!black] (6.8,2.45) -- (7.5,2.45);
\end{tikzpicture}
\caption{An illustration of \Cref{eq:row-full} for $T(x) = \{i,j\}$.  Up to the scaling factor $(1-\zeta_i^2)(1-\zeta_j^2)$, the highlighted row $x$ of $M_{\mcT}$ is a linear combination of the highlighted rows of $M$.}
\label{fig:row-combination}
\end{figure}

See~\Cref{fig:row-combination} for an illustration of~\Cref{eq:row-full}.

Using \Cref{eq:row-full} for each row of $M_{\mcT}$, we can write
\begin{equation*}
    D M_{\mcT} = F M,
\end{equation*}
where $D$ is a diagonal matrix capturing the scaling factors $\prod_{i \in T(x)} (1-\zeta_i^2)$ and each row of $F$ corresponds to the linear combination on the right-hand side of \Cref{eq:row-full}. Taking determinants, we obtain
\begin{equation*}
    \det D \det M_{\mcT} = \det F \det M.
\end{equation*}
 We then show that each of $\det D$ and $\det F$ evaluates to $1$ when $\zeta_1 = \cdots = \zeta_n = 0$, which, combined with the following elementary fact, proves the determinant equivalence lemma.
 \begin{restatable}{proposition}{minimalProp}
     \label{prop:pareto-minimal-one}
    For any polynomials $p$, $q$, if $p(0, \ldots, 0) = 1$ then $\minimal(pq) = \minimal(q)$.
 \end{restatable}

\subsubsection{Discussion: Relationship to Reimer's butterfly lemma}
\label{sec:reimer}

The antipodal routing lemma does not require $x\mapsto z(x)$ to be a permutation of $A$, but our proof produces one. We do not need this property, but we also do not know of a simpler proof that does not give a permutation.  With the permutation requirement included, condition~\ref{item:match-between} alone is equivalent to Reimer's {\sl butterfly lemma}~\cite{Rei00} (see also~\cite{KSS11}).\footnote{Without the permutation requirement, condition \ref{item:match-between} alone is easily achieved via the map $x\mapsto x^{\oplus T}$.} The name is explained by~\Cref{fig:routing}: the point $x^{\oplus T}$ is the \textsl{body} of a butterfly, and the subcubes $\cube(x,x^{\oplus T})$ and $\cube(x^{\oplus T},y)$, which meet only at the body, are its \textsl{wings}. Our proof of the antipodal routing lemma therefore strengthens the butterfly lemma by adding the coordinate-wise bound~\ref{item:match-close}.

Reimer used the butterfly lemma to prove a conjecture of van den Berg and Kesten~\cite{vdBK85}. They had shown that two monotone events occur disjointly with probability at most the product of their probabilities (the BK inequality), and conjectured that this holds for all events. Reimer's theorem is now known as the van den Berg--Kesten--Reimer (BKR) inequality. The {\sl dual} BKR inequality of Kahn, Saks, and Smyth~\cite{KSS11}, discussed in~\Cref{sec:poincare dual BKR}, is also proved using the butterfly lemma. 

Reimer's proof is also linear-algebraic, but entirely different from ours. Each point $x$ is assigned a vector in $\mathbb{F}_2^{2^n}$ that is a tensor product over the coordinates, the factor at coordinate $i$ depending on $x_i$ and on whether $i\in T$.  The lemma follows from the linear independence of these vectors. See~\cite{BCR99} for details. 

\Cref{tab:reimer} summarizes how the antipodal routing lemma relates to versions where either \ref{item:match-between} or \ref{item:match-close} is dropped.

\begin{table}[ht]
\centering
\renewcommand{\arraystretch}{1.3}
\makebox[\textwidth][c]{%
\begin{tabular}{|l|l|l|}
\hline
\textbf{Conditions} & \textbf{Statement} & \textbf{Proof} \\
\hline
\ref{item:match-between} $+$ \ref{item:match-close} & Antipodal routing lemma & determinants over $\mathbb{Z}[\zeta]$; gives a permutation \\
\ref{item:match-between} & Trivial & $z(x) = x^{\oplus T}$ \\
\ref{item:match-between} $+$ permutation & Reimer's butterfly lemma & linear algebra over $\mathbb{F}_2$ \\
\ref{item:match-close} & Faceless version (\Cref{prop:faceless matching}) & elementary; gives a permutation \\
\hline
\end{tabular}%
}
\caption{The antipodal routing lemma and related statements.}
\label{tab:reimer}
\end{table}

\section{Proof of the antipodal routing lemma}
\label{sec:matching-body}
In this section, we prove the antipodal routing lemma (\Cref{lem:matching intro}), restated here in an equivalent but more convenient form:

\begin{lemma}[Antipodal routing lemma]\label{lem:matching body}
Let $A\sse\zo^n$ and for every $x \in A$, let $T(x)\sse[n]$ be a monochromatic face of $x$. Write $\mcT\coloneqq(T(x))_{x\in A}$. Then there is $\pi = \pi_{\mcT} : A \to A$ satisfying
\begin{enumerate}[label=$(\alph*)$]
  \item\label{item:match-between-body} $\pi(x)_i \neq x_i$ for every $x \in A$ and $i \in T(x)$.
  \item\label{item:match-close-body}  For every $i\in[n]$,
  \[
    \errors_i(\pi) \leq E_i 
  \]
  where $\ds\errors_i(\pi) \coloneqq \sum_{x \in A} \Ind[\pi(x)_i = x_i]$ and $\ds E_i \coloneqq  \sum_{x \in A}\Ind[x^{\oplus i} \notin A].$
\end{enumerate}
\end{lemma}

We recall the definitions of the two  matrices $M$ and $M_{\mcT}$ from the technical overview (\Cref{sec:proof of matching lemma}):
\[    M[x,z] \coloneqq \prod_{i \in [n]} \zeta_i^{\Ind[x_i = z_i]} \quad \text{and} \quad M_{\mcT}[x,z] \coloneqq 
    \prod_{i \in [n]} \zeta_i^{\Ind[x_i = z_i]} \cdot \prod_{i \in T(x)} \Ind[x_i \neq z_i].  \]
These satisfy
\begin{align*}
    \det M &= \sum_{\text{permutations }\pi\text{ of }A} \sign(\pi) \prod_{i \in [n]} \zeta_i^{\errors_i(\pi)} \\ \det M_{\mcT} &= \sum_{\text{permutations }\pi\text{ satisfying \ref{item:match-between-body}}} \sign(\pi) \prod_{i \in [n]} \zeta_i^{\errors_i(\pi)}  .
\end{align*}
The key lemma is the following, restated for convenience.
\detequiv*
\subsection{Proof of~\texorpdfstring{\Cref{lem:matching body}}{Lemma \ref{lem:matching body}} using~\texorpdfstring{\Cref{lem:det-equiv}}{Lemma \ref{lem:det-equiv}}}
The first step is the following.
\begin{proposition}
    \label{prop:det-small}
    Every monomial in $\minimal(\det M)$ has $\zeta_i$-degree at most $E_i$ for all $i \in [n]$.
\end{proposition}
\begin{proof}
    We will show this separately for each $i \in [n]$. Fixing any such $i$, we define the monochromatic faces $\mcT^{(i)} = (T^{(i)}(x))_{x\in A}$ by
    \begin{equation*}
        T^{(i)}(x) = \begin{cases}
            \set{i}&\text{ if }x^{\oplus i} \in A\\
            \varnothing&\text{otherwise.}
        \end{cases}
    \end{equation*}
    Then $\det M_{\mcT^{(i)}}$ only sums over permutations where $\pi(x)_i \neq x_i$ for every $x$ where $x^{\oplus i} \in A$. For such permutations
\begin{equation*}
    \errors_i(\pi) = \sum_{x \in A} \Ind[\pi(x)_i = x_i]
    \leq \sum_{x \in A} \Ind[x^{\oplus i} \notin A] = E_i.
\end{equation*}
Therefore, every monomial in $\det M_{\mcT^{(i)}}$, and hence every monomial in $\minimal(\det M_{\mcT^{(i)}})$, has $\zeta_i$-degree at most $E_i$. By \Cref{lem:det-equiv}, this is also true of $\minimal(\det M)$.
\end{proof}
We next observe that:
\begin{proposition}
    \label{prop:det-survives}
    The polynomial $\minimal(\det M)$ is nonzero.
\end{proposition}
\begin{proof}
    This is equivalent to showing that $\det M$ is nonzero. For this, observe that the identity permutation $I(x) = x$ is the only permutation where $\errors_i(I) = |A|$ for all $i \in [n]$. It therefore cannot cancel with any other permutation in the determinant of $M$.
\end{proof}

The antipodal routing lemma, \Cref{lem:matching body}, follows easily from these ingredients. Combining \Cref{prop:det-small,prop:det-survives}, we have that $\minimal(\det M)$ must have a nonzero monomial with $\zeta_i$-degree at most $E_i$ for all $i\in[n]$. By \Cref{lem:det-equiv}, this is also true of $\minimal(\det M_{\mcT})$, which can only occur if there is a permutation satisfying all the constraints.

\subsection{Proof of~\texorpdfstring{\Cref{lem:det-equiv}}{Lemma \ref{lem:det-equiv}}}
We first show the following: 
\minimalProp*
\begin{proof}
    Since $p(0) = 1$, it can be written as $1 + p'$ where every monomial of $p'$ has degree at least $1$. The Pareto-minimal monomials of $pq = q + p'q$ are exactly those of $q$.
\end{proof}
The next step is the following expression linking $M$ and $M_{\mcT}$.
\begin{claim}
    \label{claim:matrix-mult}
    For any monochromatic faces $T(x)$ for every $x \in A$, let $D$ be the $A \times A$ diagonal matrix with $D[x,x] = \prod_{i \in T(x)}(1-\zeta_i^2)$ and $F$ be the $A \times A$ matrix with
    \begin{equation*}
        F[x, x^{\oplus S}] =\begin{cases} \prod_{i \in S} (-\zeta_i) &\text{if }S \subseteq T(x),\\
        0&\text{otherwise.}
    \end{cases}
\end{equation*}
    Then
    \begin{equation*}
        D M_{\mcT} = F M.
    \end{equation*}
\end{claim}
\begin{proof}
     We wish to show that, for any $x, z \in A$,
    \begin{equation*}
        (DM_{\mcT})[x,z] = (FM)[x,z].
    \end{equation*}
    Expanding matrix multiplication, this is equivalent to showing that
    \begin{equation*}
        \prod_{i \in T(x)}(1-\zeta_i^2) M_{\mcT}[x,z] = \sum_{S \subseteq T(x)} \Ind[x^{\oplus S} \in A] \cdot \paren[\bigg]{\prod_{i \in S} (-\zeta_i)} M[x^{\oplus S},z],
    \end{equation*}
    where the indicator $\Ind[x^{\oplus S} \in A]$ appears because $M$ only has rows corresponding to elements of $A$. Here, we use the assumption that $T(x)$ is a monochromatic face of $x$, meaning this condition is always satisfied. The goal is therefore to show
     \begin{equation}
        \label{eq:row-goal}
        \prod_{i \in T(x)}(1-\zeta_i^2) M_{\mcT}[x,z] = \sum_{S \subseteq T(x)} \paren[\bigg]{\prod_{i \in S} (-\zeta_i)} M[x^{\oplus S},z].
    \end{equation}
    \textbf{Warm-up: single index.} First, consider the case where $T(x) = \set{i}$. Then our desired equation is equivalent to
    \begin{equation*}
        (1-\zeta_i^2) M_{\mcT}[x,z] = M[x,z]-\zeta_i M[x^{\oplus i}, z].
    \end{equation*}
    For notational convenience, we define
    \begin{equation*}
        M^{-i}[x,z] \coloneqq \prod_{j \neq i} \zeta_j^{\Ind[x_j = z_j]}
    \end{equation*}
    to contain the part of $M[x,z]$ that does not depend on the $i$-th index. Then,
    \begin{align*}
        M[x,z] &- \zeta_i \cdot M[x^{\oplus i}, z] \\
        &= \zeta_i^{\Ind[x_i = z_i]} \cdot M^{-i}[x,z] - \zeta_i \cdot \zeta_i^{\Ind[x_i \neq z_i]} \cdot M^{-i}[x^{\oplus i},z]  \tag{Definitions of $M$, $M^{-i}$, and $x^{\oplus i}$} \\
        &= M^{-i}[x,z] \cdot \paren*{\zeta_i^{\Ind[x_i = z_i]}- \zeta_i \cdot \zeta_i^{\Ind[x_i \neq z_i]}} \tag{$M^{-i}[x, \cdot] = M^{-i}[x^{\oplus i}, \cdot]$}\\
         &= M^{-i}[x,z] \cdot \Ind[x_i \neq z_i] \cdot (1 - \zeta_i^2) \tag{Casework on $x_i = z_i$ or $x_i \neq z_i$}\\
         &= (1 - \zeta_i^2) M_{\mcT}[x,z].
    \end{align*}
    This verifies \Cref{eq:row-goal} in this simple case. Underlying the above derivation is the equality
    \begin{equation}
        \label{eq:two-by-two}
        \zeta_i^{\Ind[x_i = z_i]}- \zeta_i \cdot \zeta_i^{\Ind[x_i \neq z_i]} = \Ind[x_i \neq z_i] \cdot (1 - \zeta_i^2),
    \end{equation}
    which we will reuse shortly.

    \pparagraph{Full case.} For arbitrary $T = T(x)$, we will similarly define
    \begin{equation*}
        M^{-T}[x,z] \coloneqq \prod_{i \notin T} \zeta_i^{\Ind[x_i = z_i]}.
    \end{equation*}
    With this notation, \Cref{eq:row-goal} can be rewritten as
    \begin{equation*}
        M^{-T}[x,z]\prod_{i \in T(x)} \Ind[x_i \neq z_i](1 - \zeta_i^2)=  M^{-T}[x,z] \sum_{S \subseteq T(x)} \paren[\bigg]{\prod_{j \in S} (-\zeta_j)} \prod_{i \in T(x)} \zeta_i^{\Ind[(x^{\oplus S})_i = z_i]}.
    \end{equation*}
    It suffices to show this holds after dropping $ M^{-T}[x,z]$ from both sides:
    \begin{align*}
        \prod_{i \in T(x)} \Ind[&x_i \neq z_i](1 - \zeta_i^2) \\
        &= \prod_{i \in T(x)} \paren*{\zeta_i^{\Ind[x_i = z_i]}- \zeta_i \cdot \zeta_i^{\Ind[x_i \neq z_i]}}\tag{\Cref{eq:two-by-two}}\\
         &= \sum_{S \subseteq T(x)} \prod_{i \in S} \paren*{-\zeta_i \cdot \zeta_i^{\Ind[x_i \neq z_i]}} \cdot \prod_{i \in T(x) \setminus S} \paren*{\zeta_i^{\Ind[x_i = z_i]}}\tag{Distributive law}\\
          &= \sum_{S \subseteq T(x)} \paren[\bigg]{\prod_{i \in S}(-\zeta_i)} \cdot \paren[\bigg]{\prod_{i \in T(x)} \zeta_i^{\Ind[(x^{\oplus S})_i = z_i]}}. \tag{Collect terms by index}
    \end{align*}
    This is the equality we wished to show.
\end{proof}

We now prove the determinant equivalence lemma. We will use just two facts about the determinant:
\begin{enumerate}
    \item For any square matrices $X$ and $Y$, $\det XY = \det X \det Y$.
    \item The determinant of the identity matrix is $1$.
\end{enumerate}
\begin{proof}[Proof of \Cref{lem:det-equiv}]
    By \Cref{claim:matrix-mult},
    \begin{equation*}
        \det D \det M_{\mcT} = \det F \det M.
    \end{equation*}
    At $\zeta_1 = \cdots = \zeta_n = 0$, we have that both $D$ and $F$ become the identity matrix, which has determinant~$1$. Applying \Cref{prop:pareto-minimal-one} gives that
    \begin{equation*}
        \minimal(\det M_{\mcT}) = \minimal(\det D \det M_{\mcT}) = \minimal(\det F \det M) = \minimal(\det M). \qedhere
    \end{equation*}
\end{proof}

\section{Proof of the routing lemma}
\label{sec:routing proof}

In this section we prove the routing lemma (\Cref{lem:routing intro}) using the antipodal routing lemma (\Cref{lem:matching intro}). 

For $S\subseteq [n]$, we say that a set $Q\subseteq \zo^n$ is an {\sl $S$-subcube} if $Q=\cube(u,u^{\oplus S})$ for some $u\in\zo^{n}$. 
Each subset $S$ gives rise to $2^{n-|S|}$ different subcubes, each of which is isomorphic to $\zo^S$ and which together partition $\zo^n$. For {\sl every} $S$ and {\sl every} $S$-subcube $Q$, we are going to apply the antipodal routing lemma within that subcube.

\paragraph{Applying the antipodal routing lemma within $Q$.} Fix an $S$-subcube $Q$. The assumption of the routing lemma gives us, for every bichromatic pair $(x,y)\in A \times \bar{A}$, a monochromatic face $T(x,y)\subseteq \Delta(x, y)$ of $x$ that is contained in $A$. We use these to define, for each $x \in Q \cap A$, a monochromatic face of $x$ that is contained in $Q\cap A$:  
\[ T_Q(x)\coloneqq\begin{cases} T(x,x^{\oplus S}) & \text{if } x^{\oplus S}\notin A,\\ \varnothing & \text{otherwise.}\end{cases} \]
Since $T(x,x^{\oplus S})\sse\Delta(x,x^{\oplus S}) = S$, this monochromatic face of $x$ is indeed contained in $Q\cap A$. In the second case, the empty face is trivially monochromatic. 

We can therefore apply the antipodal routing lemma to the set $A\cap Q$ in $Q\cong\zo^S$, with faces $T_Q(x)$. In $Q$, the antipode of $x$ is $x^{\oplus S}$. Thus, for all $x\in Q\cap A$ the antipodal routing lemma gives a point $z_Q(x)\in Q\cap A$ such that 
\begin{enumerate}[label=$(\alph*^Q)$]
  \item\label{item:Q-between} $z_Q(x)\in\cube\big(x^{\oplus T_Q(x)},\,x^{\oplus S}\big)$, and
  \item\label{item:Q-close} for every $i\in S$, \ $\Prx_{\bx\sim Q}\big[\bx\in A,\ z_Q(\bx)_i\neq(\bx^{\oplus S})_i\big]\le\lfrac12\cdot\Inf_i^Q(A)$
\end{enumerate}
where $\Inf_i^Q(A)\coloneqq\Prx_{\bx\sim Q}[\Ind[\bx\in A]\neq\Ind[\bx^{\oplus i}\in A]]$ is the influence of the $i$-th variable within the subcube $Q$. 

Finally, we are ready to define $z(x,y,T)$ as required by the routing lemma. We associate each bichromatic pair $(x, y)$ with the subcube in which they are antipodal. This corresponds exactly to the $\Delta(x,y)$-subcube $Q=\cube(x,y)$. We let 
\[ z(x,y,T)\coloneqq z_Q(x) \qquad\text{for $Q=\cube(x,y)$}. \]

Note that $z(x,y,T)\in A$ as required. The remainder of the proof will show that with this choice of $z$, items \ref{item:between} and \ref{item:close} of the routing lemma hold. 

\paragraph{Item \ref{item:between} holds.} Since  $(x,y)\in A \times \bar{A}$ is a bichromatic pair, the first case in the definition of $T_Q$ applies, and so $T_Q(x)=T(x,y)$. Then, with our choice of $z(x,y,T)$, \ref{item:Q-between} becomes $z(x,y,T)\in  \cube\big(x^{\oplus T(x,y)},\,y\big)$ as desired. This comes simply from plugging in $S=\Delta(x,y)$ and noting that $x^{\oplus \Delta(x,y)}=y$.

\paragraph{Item \ref{item:close} holds.} A uniform random pair $(\bx,\by)$ can be sampled as follows: draw $\bx\sim\zo^n$ and $\bS\sse[n]$ uniformly and independently, and set $\by\coloneqq\bx^{\oplus\bS}$. Then $\bS = \Delta(\bx,\by)$, and $\by$ is the antipode of $\bx$ in the $\bS$-subcube containing $\bx$. Each coordinate lies in $\bS$ with probability $\frac12$. We can therefore rewrite the inequality we wish to prove as:
\begin{align*}
     \Prx_{\bx,\bS}\big[\bx\in A,\ \bx^{\oplus\bS} \notin A,\ \bz_i\neq(\bx^{\oplus\bS})_i\big] \;\leq\; \lfrac14\cdot\Inf_i(A).
\end{align*}
Let $E$ be the event $\{\bx\in A,\ \bx^{\oplus\bS} \notin A,\ \bz_i\neq(\bx^{\oplus\bS})_i\}$ whose probability we wish to bound. By the law of total probability,
\begin{align}
    \Pr[E] &= \Pr[i\in\bS]\cdot\Pr\big[E \,\big|\, i\in\bS\big] \;+\; \Pr[i\notin\bS]\cdot\Pr\big[E \,\big|\, i\notin\bS\big] \nonumber \\
    &= \lfrac12 \cdot \Pr\big[E \,\big|\, i\in\bS\big]. \label{eq:PrE}
\end{align}
The second equality follows from the fact that if $i\notin\bS$, then $\bz$ and $\by=\bx^{\oplus\bS}$ both lie in the $\bS$-subcube containing $\bx$, whose points all agree outside $\bS$. Hence $\bz_i=\by_i$, and so $\Pr[E\mid i\notin\bS]=0$.

It remains to show that $\Pr[E\mid i\in\bS]\le\frac12\Inf_i(A)$. Fix any $S\ni i$. Given $\bS=S$, the point $\bx$ is uniform on $\zo^n$, which is the same as drawing a uniform $S$-subcube $\bQ$ and then a uniform $\bx\sim\bQ$. Conditioned on $\bQ=Q$, we have $\bz=z_Q(\bx)$ and $\by=\bx^{\oplus S}$, so dropping the condition $\by\notin A$ and applying~\ref{item:Q-close},
\[
  \Pr\big[E\,\big|\,\bS=S,\ \bQ=Q\big]\;\le\;\Prx_{\bx\sim Q}\big[\bx\in A,\ z_Q(\bx)_i\neq(\bx^{\oplus S})_i\big]\;\le\;\lfrac12\cdot\Inf_i^{Q}(A).
\]
Since $i\in S$, every edge $\{x,x^{\oplus i}\}$ lies inside a single $S$-subcube, and so averaging over $\bQ$ gives $\Ex_{\bQ}\big[\Inf_i^{\bQ}(A)\big]=\Inf_i(A)$. Hence $\Pr[E\mid\bS=S]\le\frac12\cdot \Inf_i(A)$ for every $S\ni i$, and therefore $\Pr[E\mid i\in\bS]\le\frac12\cdot \Inf_i(A)$. Combining this with~\Cref{eq:PrE} completes the proof.

\section{Proof of the weighted all-pairs inequality}

We prove the weighted all-pairs inequality (\Cref{lem:weighted}) assuming the routing lemma (\Cref{lem:routing intro}). The argument is identical to the unweighted case sketched in the technical overview (\Cref{sec:overview-routing-to-weighted}), suitably modified to take the weights into account. 

Let $A \coloneqq f^{-1}(1)$. For any pair $x\in A$ and $y\notin A$, let $\tau = \overlap_{w,\Lambda}(x,y)$ and $T = \{ i \in \Delta(x,y) \colon w_i\lambda_y(i) > \tau\}$ be the heavy coordinates of $\lambda_y$ weighted according to $w$.  By the same Markov argument as given in the technical overview, our choice of $\tau$ ensures $\lambda_x(T)<1$, which in turn ensures that $T$ is a monochromatic face of $x$.

Applying the routing lemma with this choice of monochromatic faces and taking a $w$-weighted sum over $i\in [n]$ in condition \ref{item:close}, we get  
\begin{equation}\label{eq:close-summed-weighted}  \Ex_{\bx,\by}\big[\Ind[\bx \in A, \by\notin A]\cdot \dist_w(\bz,\by)\big] \le \frac1{4} \sum_{i=1}^n\frac{\Inf_i(A)}{w_i}, 
\end{equation}
where $\dist_w(z,y)\coloneqq \sum_{i\in\Delta(z,y)} \frac{1}{w_i}$ is the $w$-weighted Hamming distance between $z$ and $y$.  \smallskip

We lower bound the LHS of~\Cref{eq:close-summed-weighted} for each $(x,y)$ pair. By condition \ref{item:between} of the routing lemma, the point $z$ satisfies $\Delta(z,y)\sse \Delta(x,y)\setminus T$, and so
\[ \lambda_y(i) \le  \frac{\tau}{w_i}\quad \text{for all $i\in \Delta(z,y)$.}\] 
And yet $z\in A$, so these coordinates must furthermore satisfy $\sum_{i\in \Delta(z,y)} \lambda_y(i) \ge 1$. Putting these two observations together, \[ 1\le \sum_{i\in \Delta(z,y)} \lambda_y(i) \le \sum_{i\in \Delta(z,y)}\frac{\tau}{w_i} = \tau\cdot \dist_w(z,y).\]
By our choice of $\tau$, we have shown that
\[ \dist_w(z,y) \ge \frac1{\overlap_{w,\Lambda}(x,y)}.\]
Averaging this inequality over $\bx,\by$ and combining it with~\Cref{eq:close-summed-weighted} yields the weighted all-pairs inequality.

\section{Proof of the certificate-influence inequality}\label{sec:prices-to-key}

We prove the certificate-influence inequality (\Cref{lem:key}) using the weighted all-pairs inequality (\Cref{lem:weighted}). Throughout this proof, let $\bx$ be uniform on $f^{-1}(1)$ and $\by$ independently uniform on $f^{-1}(0)$.  The certificate-influence inequality can be written as:  
\begin{equation}
\label{eq:goal-with-var}
4\underbrace{\Pr[f=1]\Pr[f=0]}_{= \Var(f)} \cdot \Ex_{\bx}\left[\frac{1}{\Lone{\lambda_{\bx}}}\right]
  \cdot
  \Ex_{\by}\left[\frac{1}{\Lone{\lambda_{\by}}}\right]
  \;\leq\;  \max_{i\in[n]} \Inf_i(f) \eqqcolon M.
\end{equation}
In this notation, the weighted all-pairs inequality is
\begin{equation}
    \label{eq:maxinf-weighted}
    4\Var(f)\cdot \Ex_{\bx,\by}\bracket*{\frac{1}{\overlap_{w, \Lambda}(\bx,\by)}} \leq \sum_{i \in [n]} \frac{\Inf_i(f)}{w_i} \leq \sum_{i \in [n]} \frac{1}{w_i} \cdot M.
\end{equation}
We will use Cauchy--Schwarz to upper bound the left side of our goal with two terms, one of which we can bound using \Cref{lem:weighted}.
\begin{align*}
    \bigg(\Ex_{\bx}\left[\lfrac{1}{\Lone{\lambda_{\bx}}}\right]
  \cdot
  \Ex_{\by}\left[\lfrac{1}{\Lone{\lambda_{\by}}}\right]\bigg)^2 
 &= \Ex_{\bx,\by}\bracket*{\lfrac{1}{\lone{\lambda_{\bx}}\cdot {\lone{\lambda_{\by}}}}}^2\tag{Independence}\\
  &= \Ex_{\bx,\by}\bracket*{\paren*{\lfrac{\sqrt{\overlap_{w, \Lambda}(\bx,\by)}}{\lone{\lambda_{\bx}}\cdot {\lone{\lambda_{\by}}}}} \cdot \paren*{\lfrac{1}{\sqrt{\overlap_{w, \Lambda}(\bx,\by)}}}}^2\\
 &\leq\Ex_{\bx,\by}\bracket*{\paren*{\lfrac{\sqrt{\overlap_{w, \Lambda}(\bx,\by)}}{\lone{\lambda_{\bx}}\cdot {\lone{\lambda_{\by}}}}}^2} \cdot \Ex_{\bx,\by}\bracket*{ \paren*{\lfrac{1}{\sqrt{\overlap_{w, \Lambda}(\bx,\by)}}}^2}\tag{Cauchy--Schwarz}\\
   &\leq\Ex_{\bx,\by}\bracket*{\paren*{\lfrac{\sqrt{\overlap_{w, \Lambda}(\bx,\by)}}{\lone{\lambda_{\bx}}\cdot {\lone{\lambda_{\by}}}}}^2} \cdot \sum_{i \in [n]} \frac{1}{w_i} \cdot \frac{M}{4\Var(f)} .\tag{\Cref{eq:maxinf-weighted}}
\end{align*}
We now upper bound the first term in the above expression:
\begin{align*}
    \Ex_{\bx,\by}\bracket*{\paren*{\lfrac{\sqrt{\overlap_{w, \Lambda}(\bx,\by)}}{\lone{\lambda_{\bx}}\cdot {\lone{\lambda_{\by}}}}}^2} &= \Ex_{\bx,\by}\bracket*{\frac{\sum_{i \in \Delta(\bx,\by)} w_i\lambda_{\bx}(i) \lambda_{\by}(i)}{\lone{\lambda_{\bx}}^2\cdot {\lone{\lambda_{\by}}^2}}}\tag{Definition of $\overlap$}\\
    &\leq \Ex_{\bx,\by}\bracket*{\frac{\sum_{i \in [n]} w_i\lambda_{\bx}(i) \lambda_{\by}(i)}{\lone{\lambda_{\bx}}^2\cdot {\lone{\lambda_{\by}}^2}}}\tag{Sum over all indices}\\
    &= \sum_{i \in [n]} w_i \cdot \Ex_{\bx,\by}\bracket*{\frac{\lambda_{\bx}(i) \lambda_{\by}(i)}{\lone{\lambda_{\bx}}^2\cdot {\lone{\lambda_{\by}}^2}}}\tag{Linearity of expectation}\\
    &= \sum_{i \in [n]} w_i \cdot \Ex_{\bx}\bracket*{\frac{\lambda_{\bx}(i) }{\lone{\lambda_{\bx}}^2}}\cdot \Ex_{\by}\bracket*{\frac{\lambda_{\by}(i) }{\lone{\lambda_{\by}}^2}} .\tag{Independence of $\bx$ and $\by$}
\end{align*}
We will choose $w_i = \paren*{\Ex_{\bx}\bracket*{\frac{\lambda_{\bx}(i) }{\lone{\lambda_{\bx}}^2}}}^{-1}$ to cancel the first term.\footnote{To avoid dividing by $0$, we formally set $w_i = \paren*{\eps + \Ex_{\bx}\bracket*{\frac{\lambda_{\bx}(i) }{\lone{\lambda_{\bx}}^2}}}^{-1}$ and take the limit as $\eps \to 0$. The resulting proof is identical up to bookkeeping.} Combining these expressions, we then have that
\begin{align*}
    \paren*{\Ex_{\bx}\left[\frac{1}{\Lone{\lambda_{\bx}}}\right]
  \cdot
  \Ex_{\by}\left[\frac{1}{\Lone{\lambda_{\by}}}\right]}^2 &\leq \Ex_{\bx,\by}\bracket*{\paren*{\lfrac{\sqrt{\overlap_{w, \Lambda}(\bx,\by)}}{\lone{\lambda_{\bx}}\cdot {\lone{\lambda_{\by}}}}}^2} \cdot \sum_{i \in [n]} \frac{1}{w_i} \cdot \frac{M}{4\Var(f)}  \\
  &\leq\sum_{i \in [n]}\Ex_{\by}\bracket*{\frac{\lambda_{\by}(i) }{\lone{\lambda_{\by}}^2}} \cdot \sum_{i \in [n]} \Ex_{\bx}\bracket*{\frac{\lambda_{\bx}(i) }{\lone{\lambda_{\bx}}^2}} \cdot \frac{M}{4\Var(f)}\\
  &= \Ex_{\by}\bracket*{\frac{\lone{\lambda_{\by}}}{\lone{\lambda_{\by}}^2}} \cdot \Ex_{\bx}\bracket*{\frac{\lone{\lambda_{\bx}} }{\lone{\lambda_{\bx}}^2}} \cdot \frac{M}{4\Var(f)} \tag{Linearity of expectation} \\
  &= \Ex_{\by}\bracket*{\frac{1}{\lone{\lambda_{\by}}}} \cdot \Ex_{\bx}\bracket*{\frac{1 }{\lone{\lambda_{\bx}}}} \cdot \frac{M}{4\Var(f)}.
\end{align*}
Rearranging gives \Cref{eq:goal-with-var}, which is what we wanted to show.

\section{Proof of the AA conjecture}
\label{sec:AA proof}

We now prove the AA conjecture (\Cref{thm:intro-AA}) using the certificate-influence inequality (\Cref{lem:key}). We in fact prove a more general statement that gives a lower bound on the maximum influence of an arbitrary function $p\colon\zo^n\to\R$ in terms of its fractional certificate complexity, with no assumptions on degree or boundedness:

\begin{theorem}[AA-like statement for arbitrary functions]\label{thm:FC sigma}
Let $p\colon\zo^n\to\R$ be nonconstant,  $\mu\coloneqq\E[p(\bx)]$, and $\sigma\coloneqq\E[|p(\bx)-\mu|]$. Then
\[
  \max_{i\in[n]}\Inf_i(p) \;\geq\; \Omega\left(\frac{\sigma^{8}}{\FC(p)^6}\right).
\]
\end{theorem}

\begin{proof}[Proof of~\Cref{thm:intro-AA} (the AA conjecture) assuming~\Cref{thm:FC sigma}]
Let $p : \zo^n \to [0,1]$ be a bounded degree-$d$ polynomial. By \Cref{fact:ABD}, $\FC(p) = O(d^2)$ (this uses both assumptions of low degree and boundedness). Since $p$ takes values in $[0,1]$, we have that $|p-\mu|\le1$, hence $|p-\mu|\geq(p-\mu)^2$, and so $\sigma\geq\Var(p)$ (this uses only boundedness). \Cref{thm:intro-AA} now follows from \Cref{thm:FC sigma}.
\end{proof}

\subsection{Proof of~\texorpdfstring{\Cref{thm:FC sigma}}{Theorem \ref{thm:FC sigma}}}
Fix a fractional certificate family $\Lambda = (\lambda_x)_{x\in\zo^n}$ for $p$ of cost $\FC(p)$. For a threshold $t\in\R$, let $p_t(x)\coloneqq\Ind[p(x)>t]$. 

We first claim that for each $x\in\zo^n$, the scaled vector \[ \frac{\lambda_x}{|p(x)-t|}\]  
is a fractional certificate for $p_t$ at $x$.\footnote{Strictly speaking, these certificates are undefined when $p(x)= t$. However, since there are only finitely many $x$'s and we will be choosing $t$ uniformly from an interval, this bad event happens with probability $0$.} Indeed, if $p_t(x)\neq p_t(y)$, then $p(x)$ and $p(y)$ lie on opposite sides of $t$, and so
\[
  \lambda_x(\Delta(x,y)) \;\geq\; |p(x)-p(y)| \;\geq\; |p(x)-t|.
\]
Hence, whenever $p_t$ is nonconstant, we can apply the certificate-influence inequality to get
\begin{equation} \max_{i\in [n]}\Inf_i(p_t) 
\ge 4 \cdot \frac{\E\big[ (p(\bx)-t)_{+}\big] \Ex\big[(t-p(\by))_{+}\big]}{\FC(p)^2}.
\label{eq:p_t max inf}
\end{equation}
Since $\E[(p(\bx)-\mu)_+] = \E[(\mu-p(\bx))_+]$ and they sum to $\sigma$, each equals $\sigma/2$. The maps $t\mapsto\E[(p(\bx)-t)_+]$ and $t\mapsto\E[(t-p(\bx))_+]$ are $1$-Lipschitz, so for $t \in [\mu-\frac{\sigma}{4},\mu+\frac{\sigma}{4}]$ both are at least $\sigma/4$. In particular, for any such $t$ we have that $p_t$ is nonconstant and the numerator on the RHS of~\Cref{eq:p_t max inf} is $\Omega(\sigma^2)$.

Next, we relate $\Inf_i(p_t)$ and $\Inf_i(p)$. We first write:
\[
\Inf_i(p_t) = 2\, \Pr\big[p(\bx^{\oplus i}) \le t < p(\bx)\big].
\]
Fix $\theta>0$. If $p(x^{\oplus i})\le t < p(x)$, then either $p(x)$ lies in the window $(t,t+\theta\,]$, or the jump $p(x)-p(x^{\oplus i})$ exceeds $\theta$. The probability of the latter is at most
\[  \Pr\big[p(\bx)-p(\bx^{\oplus i})>\theta\big] = \tfrac{1}{2} \Pr\big[|p(\bx)-p(\bx^{\oplus i})|>\theta\big] \le \frac{\Inf_i(p)}{2\theta^2}.%\frac{\Infone_i(p)}{2\theta},
\]
where $\ds \Infone_i(p) \coloneqq \Ex_{\bx}[|p(\bx) - p(\bx^{\oplus i})|]$ is the $L_1$ version of influence. 
The equality is by symmetry and the inequality is Markov's inequality applied to the quantity $(p(\bx)-p(\bx^{\oplus i}))^2$. Therefore, 
\begin{equation*}
  \Inf_i(p_t) \le 2\,\Pr\big[t<p(\bx)\leq t+\theta\big] + \frac{\Inf_i(p)}{\theta^2}.
\end{equation*}

Combining~\Cref{eq:p_t max inf} with this window-jump argument, we have shown that for every $t\in[\mu-\frac{\sigma}{4},\mu+\frac{\sigma}{4}]$ and $\theta>0$,  
\begin{equation} \frac{M}{\theta^2} + 2\,\Pr\big[t<p(\bx)\leq t+\theta\big] \ge \max_{i\in [n]}\Inf_i(p_t) \ge \Omega\left(\frac{\sigma^2}{\FC(p)^2}\right),\label{eq:AA with theta}
\end{equation}
where $M\coloneqq \max_{i\in [n]}\Inf_i(p)$.  Now let $\bt$ be uniform on $[\mu-\frac{\sigma}{4},\mu+\frac{\sigma}{4}]$, an interval of length $\sigma/2$, and note that
\[ \Ex_{\bt}\Big[\Prx_{\bx}\big[\bt<p(\bx)\leq\bt+\theta\big]\Big] \;=\; \Ex_{\bx}\Big[\Prx_{\bt}\big[p(\bx)-\theta\leq\bt<p(\bx)\big]\Big] \;\leq\; \frac{\theta}{\sigma/2} \;=\; \frac{2\theta}{\sigma}.
\]
Averaging~\Cref{eq:AA with theta} over $\bt$ therefore gives, for every $\theta>0$,
\[ 
\frac{M}{\theta^2} + \frac{4\theta}{\sigma}\ge \Omega\left(\frac{\sigma^2}{\FC(p)^2}\right).
\]
Finally, choosing $\theta = (\sigma M)^{1/3}$ and rearranging, we get: 
\begin{equation*}
    \max_{i\in[n]}\Inf_i(p) \ge \Omega\left(\frac{\sigma^8}{\FC(p)^6}\right). 
\end{equation*}

\section{Acknowledgments}

We thank PSI, and in particular David Turturean and Jim Halverson, for sharing the proof with us and suggesting that we write this exposition. We also thank Scott Aaronson for putting the two teams in touch. 

Guy, Carmen, and Li-Yang are supported by NSF awards 1942123, 2211237, 2224246, a Sloan Research
Fellowship, and a Google Research Scholar Award. Jordan is supported by  NSF awards 2016245 and 2440805 and by the Air Force Office of Scientific Research under grant agreement FA9550-21-1-0392.  Guy is also supported by a Norbert Wiener Postdoctoral Fellowship and Omer Reingold’s Simons Foundation Investigator Award. Jordan is also supported by the Shoucheng Zhang Graduate Fellowship. Carmen is also supported by a Stanford
Computer Science Distinguished Fellowship, an NSF GRFP, Omer Reingold’s Simons Foundation Investigator Award, and a Jump Trading PhD Fellowship. 
Any opinions, findings, and 
conclusions or recommendations expressed in this material are those of the authors and do not reflect 
the views of Jump Trading.

\bibliography{ref}

\newcommand{\etalchar}[1]{$^{#1}$}
\begin{thebibliography}{GMNR26}

\bibitem[AA14]{AA14}
Scott Aaronson and Andris Ambainis.
\newblock The need for structure in quantum speedups.
\newblock {\em Theory of Computing}, 10(6):133--166, 2014.
\newblock Preliminary version in ICS 2011.

\bibitem[Aar05]{Aar05}
Scott Aaronson.
\newblock Ten semi-grand challenges for quantum computing theory, 2005.
\newblock \url{https://www.scottaaronson.com/writings/qchallenge.html}.

\bibitem[Aar08a]{Aar08talk}
Scott Aaronson.
\newblock How to solve longstanding open problems in quantum computing using only {F}ourier analysis, 2008.
\newblock Lecture at the Banff International Research Station. \url{https://www.scottaaronson.com/talks/openqc.ppt}.

\bibitem[Aar08b]{Aar08}
Scott Aaronson.
\newblock Quantum certificate complexity.
\newblock {\em Journal of Computer and System Sciences}, 74(3):313--322, 2008.
\newblock Preliminary version in CCC 2003.

\bibitem[ABD26]{ABD26}
Avantika Agarwal and Shalev Ben-David.
\newblock Oracle separations for the quantum-classical polynomial hierarchy.
\newblock In {\em Proceedings of the 17th Innovations in Theoretical Computer Science Conference (ITCS 2026)}, pages 2:1--2:22, 2026.

\bibitem[ABDK21]{ABK21}
Anurag Anshu, Shalev Ben-David, and Srijita Kundu.
\newblock On query-to-communication lifting for adversary bounds.
\newblock In {\em Proceedings of the 36th Computational Complexity Conference (CCC 2021)}, pages 30:1--30:39, 2021.

\bibitem[ABP19]{ABP19}
Srinivasan Arunachalam, Jop Bri\"et, and Carlos Palazuelos.
\newblock Quantum query algorithms are completely bounded.
\newblock {\em SIAM Journal on Computing}, 48(3):903--925, 2019.
\newblock Preliminary version in ITCS 2018.

\bibitem[BBBV97]{BBBV97}
Charles Bennett, Ethan Bernstein, Gilles Brassard, and Umesh Vazirani.
\newblock Strengths and weaknesses of quantum computing.
\newblock {\em SIAM Journal on Computing}, 26(5):1510--1523, 1997.

\bibitem[BBC{\etalchar{+}}01]{BBCMW01}
Robert Beals, Harry Buhrman, Richard Cleve, Michele Mosca, and Ronald de~Wolf.
\newblock Quantum lower bounds by polynomials.
\newblock {\em Journal of the ACM}, 48(4):778--797, 2001.
\newblock Preliminary version in FOCS 1998.

\bibitem[BCR99]{BCR99}
Christian Borgs, Jennifer Chayes, and Dana Randall.
\newblock The van den {B}erg--{K}esten--{R}eimer inequality: A review.
\newblock In {\em Perplexing Problems in Probability: Festschrift in Honor of {H}arry {K}esten}, volume~44 of {\em Progress in Probability}, pages 159--173. Birkh{\"a}user, 1999.

\bibitem[BDST26]{BDST26}
Guy Blanc, Jordan Docter, Carmen Strassle, and Li-Yang Tan.
\newblock Quantum speedups require structure or depth.
\newblock In {\em Proceedings of the 67th Annual Symposium on Foundations of Computer Science (FOCS 2026)}, 2026.

\bibitem[BG81]{BG81}
Charles Bennett and John Gill.
\newblock Relative to a random oracle {$A$}, {$\mathsf{P}^A \neq \mathsf{NP}^A \neq \mathsf{co}\text{-}\mathsf{NP}^A$} with probability 1.
\newblock {\em SIAM Journal on Computing}, 10(1):96--113, 1981.

\bibitem[Bha25]{Bha25}
Sreejata~Kishor Bhattacharya.
\newblock Random restrictions of bounded low degree polynomials are juntas.
\newblock In {\em Proceedings of the 16th Innovations in Theoretical Computer Science Conference (ITCS 2025)}, pages 17:1--17:21, 2025.

\bibitem[BSdW22]{BSdW22}
Nikhil Bansal, Makrand Sinha, and Ronald de~Wolf.
\newblock Influence in completely bounded block-multilinear forms and classical simulation of quantum algorithms.
\newblock In {\em Proceedings of the 37th Computational Complexity Conference (CCC 2022)}, pages 28:1--28:21, 2022.

\bibitem[BV97]{BV97}
Ethan Bernstein and Umesh Vazirani.
\newblock Quantum complexity theory.
\newblock {\em SIAM Journal on Computing}, 26(5):1411--1473, 1997.
\newblock Preliminary version in STOC 1993.

\bibitem[CCG{\etalchar{+}}94]{CCGHHRR94}
Richard Chang, Benny Chor, Oded Goldreich, Juris Hartmanis, Johan H{\aa}stad, Desh Ranjan, and Pankaj Rohatgi.
\newblock The random oracle hypothesis is false.
\newblock {\em Journal of Computer and System Sciences}, 49(1):24--39, 1994.

\bibitem[DFKO07]{DFKO07}
Irit Dinur, Ehud Friedgut, Guy Kindler, and Ryan O'Donnell.
\newblock On the {F}ourier tails of bounded functions over the discrete cube.
\newblock {\em Israel Journal of Mathematics}, 160:389--412, 2007.
\newblock Preliminary version in STOC 2006.

\bibitem[DMP19]{DMP19}
Andreas Defant, Mieczys{\l}aw Masty{\l}o, and Antonio P\'erez.
\newblock On the {F}ourier spectrum of functions on {B}oolean cubes.
\newblock {\em Mathematische Annalen}, 374(1):653--680, 2019.

\bibitem[{Esc}24]{EG24}
Francisco {Escudero Guti\'errez}.
\newblock Influences of {F}ourier completely bounded polynomials and classical simulation of quantum algorithms.
\newblock {\em Chicago Journal of Theoretical Computer Science}, 2024(2), 2024.

\bibitem[ESP26]{EPS26}
Francisco {Escudero Guti\'errez}, Miquel Saucedo, and Carlos Palazuelos.
\newblock Optimal inequalities for completely bounded polynomials and the limitations of quantum query algorithms, 2026.
\newblock arXiv:2609.05201.

\bibitem[For21]{For21}
Lance Fortnow.
\newblock Worlds to die harder for: Open oracle questions for the 21st century.
\newblock {\em ACM SIGACT News}, 52(3):26--36, 2021.

\bibitem[FR99]{FR99}
Lance Fortnow and John Rogers.
\newblock Complexity limitations on quantum computation.
\newblock {\em Journal of Computer and System Sciences}, 59(2):240--252, 1999.
\newblock Preliminary version in CCC 1998.

\bibitem[GMNR26]{GMNR26}
Uma Girish, Kunal Mittal, Barak Nehoran, and Ran Raz.
\newblock Distributional variants of the {A}aronson--{A}mbainis conjecture, 2026.
\newblock arXiv:2609.35327.

\bibitem[Gro96]{Gro96}
Lov Grover.
\newblock A fast quantum mechanical algorithm for database search.
\newblock In {\em Proceedings of the 28th Annual Symposium on Theory of Computing (STOC 1996)}, pages 212--219, 1996.

\bibitem[GSS16]{GSS16}
Justin Gilmer, Michael Saks, and Srikanth Srinivasan.
\newblock Composition limits and separating examples for some {B}oolean function complexity measures.
\newblock {\em Combinatorica}, 36(3):265--311, 2016.
\newblock Preliminary version in CCC 2013.

\bibitem[Har85]{Har85}
Juris Hartmanis.
\newblock Solvable problems with conflicting relativizations.
\newblock {\em Bulletin of the European Association for Theoretical Computer Science}, 27:40--49, 1985.

\bibitem[Iva26]{Iva26}
Paata Ivanisvili.
\newblock Polynomial growth of {B}ohnenblust--{H}ille constants on the {H}amming cube, 2026.
\newblock arXiv:2609.12427.

\bibitem[KSS11]{KSS11}
Jeff Kahn, Michael Saks, and Clifford Smyth.
\newblock The dual {BKR} inequality and {R}udich's conjecture.
\newblock {\em Combinatorics, Probability and Computing}, 20(2):257--266, 2011.
\newblock Preliminary version in CCC 2000.

\bibitem[KT16]{KT16}
Raghav Kulkarni and Avishay Tal.
\newblock On fractional block sensitivity.
\newblock {\em Chicago Journal of Theoretical Computer Science}, 2016(8), 2016.

\bibitem[Kur82]{Kur82}
Stuart Kurtz.
\newblock On the random oracle hypothesis.
\newblock In {\em Proceedings of the 14th Annual Symposium on Theory of Computing (STOC 1982)}, pages 224--230, 1982.

\bibitem[LM26]{LM26}
Qipeng Liu and Saachi Mutreja.
\newblock Parallel quantum advantage with limited adaptivity requires structure, 2026.
\newblock arXiv:2608.20297.

\bibitem[LZ23]{LZ23}
Shachar Lovett and Jiapeng Zhang.
\newblock Fractional certificates for bounded functions.
\newblock In {\em Proceedings of the 14th Innovations in Theoretical Computer Science Conference (ITCS 2023)}, pages 84:1--84:13, 2023.

\bibitem[Mon12]{Mon12}
Ashley Montanaro.
\newblock Some applications of hypercontractive inequalities in quantum information theory.
\newblock {\em Journal of Mathematical Physics}, 53(12):122206, 2012.

\bibitem[Nis91]{Nis91}
Noam Nisan.
\newblock {CREW} {PRAM}s and decision trees.
\newblock {\em SIAM Journal on Computing}, 20(6):999--1007, 1991.
\newblock Preliminary version in STOC 1989.

\bibitem[NS94]{NS94}
Noam Nisan and Mario Szegedy.
\newblock On the degree of {B}oolean functions as real polynomials.
\newblock {\em Computational Complexity}, 4(4):301--313, 1994.
\newblock Preliminary version in STOC 1992.

\bibitem[OSSS05]{OSSS05}
Ryan O'Donnell, Michael Saks, Oded Schramm, and Rocco Servedio.
\newblock Every decision tree has an influential variable.
\newblock In {\em Proceedings of the 46th Annual Symposium on Foundations of Computer Science (FOCS 2005)}, pages 31--39, 2005.

\bibitem[OZ16]{OZ16}
Ryan O'Donnell and Yu~Zhao.
\newblock Polynomial bounds for decoupling, with applications.
\newblock In {\em Proceedings of the 31st Conference on Computational Complexity (CCC 2016)}, pages 24:1--24:18, 2016.

\bibitem[Rei00]{Rei00}
David Reimer.
\newblock Proof of the van den {B}erg--{K}esten conjecture.
\newblock {\em Combinatorics, Probability and Computing}, 9(1):27--32, 2000.

\bibitem[Rud88]{Rud88}
Steven Rudich.
\newblock {\em Limits on the Provable Consequences of One-Way Functions}.
\newblock PhD thesis, University of California, Berkeley, 1988.

\bibitem[Sho97]{Sho97}
Peter Shor.
\newblock Polynomial-time algorithms for prime factorization and discrete logarithms on a quantum computer.
\newblock {\em SIAM Journal on Computing}, 26(5):1484--1509, 1997.
\newblock Preliminary version in FOCS 1994.

\bibitem[Sim97]{Sim97}
Daniel Simon.
\newblock On the power of quantum computation.
\newblock {\em SIAM Journal on Computing}, 26(5):1474--1483, 1997.
\newblock Preliminary version in FOCS 1994.

\bibitem[Smy01]{Smy01}
Clifford Smyth.
\newblock {\em The Conjectures of {R}udich, {T}ardos, and {K}usner}.
\newblock PhD thesis, Rutgers, The State University of New Jersey, 2001.

\bibitem[Smy02]{Smy02}
Clifford Smyth.
\newblock Reimer's inequality and {T}ardos' conjecture.
\newblock In {\em Proceedings of the 34th Annual Symposium on Theory of Computing (STOC 2002)}, pages 218--221, 2002.

\bibitem[STV26]{STV26}
Joseph Slote, Chun-Kai Tseng, and Alexander Volberg.
\newblock An $m^{2.943}$ {B}ohnenblust--{H}ille bound on the {B}oolean cube, 2026.
\newblock arXiv:2609.21144.

\bibitem[Tal96]{Tal96}
Michel Talagrand.
\newblock A new look at independence.
\newblock {\em Annals of Probability}, 24(1):1--34, 1996.

\bibitem[Tal13]{Tal13}
Avishay Tal.
\newblock Properties and applications of {B}oolean function composition.
\newblock In {\em Proceedings of the 4th Conference on Innovations in Theoretical Computer Science (ITCS 2013)}, pages 441--454, 2013.

\bibitem[Tar89]{Tar89}
G\'abor Tardos.
\newblock Query complexity, or why is it difficult to separate {$\mathsf{NP}^A \cap \mathsf{coNP}^A$} from {$\mathsf{P}^A$} by random oracles {$A$}?
\newblock {\em Combinatorica}, 9(4):385--392, 1989.

\bibitem[Tur26]{Tur26}
David Turturean.
\newblock {A Proof of the Aaronson--Ambainis Conjecture}, November 2026.
\newblock Manuscript. \url{https://psi.inc/research/aaronson-ambainis/}.

\bibitem[vdBK85]{vdBK85}
Jacob van~den Berg and Harry Kesten.
\newblock Inequalities with applications to percolation and reliability.
\newblock {\em Journal of Applied Probability}, 22(3):556--569, 1985.

\bibitem[YZ24]{YZ24}
Takashi Yamakawa and Mark Zhandry.
\newblock Verifiable quantum advantage without structure.
\newblock {\em Journal of the ACM}, 71(3):20, 2024.
\newblock Preliminary version in FOCS 2022.

\end{thebibliography}
\bibliographystyle{alpha}

%\appendix 

%\newpage 

\end{document}